\documentclass[12pt, a4paper]{article}

\pdfoutput=1

\usepackage{amsmath}
\allowdisplaybreaks
\usepackage{amsfonts}
\usepackage{amssymb}
\usepackage{bbm}
\usepackage{verbatim}
\usepackage{dsfont}
\usepackage{booktabs}
\usepackage{slashed}

\usepackage{epsfig}
\usepackage{color}
\usepackage[table]{xcolor}
\usepackage{graphicx}
\usepackage[]{caption}
\usepackage[listofformat=empty,subrefformat=empty]{subfig}

\usepackage{float}
\usepackage{overpic}

\usepackage{enumerate}
\usepackage{hhline}
\usepackage{multirow}

\usepackage{cite}
\usepackage{xspace}
\usepackage{setspace}

\usepackage[pdftitle={Effective field theory for magnetic compactifications},
  pdfauthor={},
  pdfsubject={},
  bookmarksopen, bookmarksnumbered, bookmarksopenlevel=2, colorlinks=false, linkcolor=blue, citecolor=blue, urlcolor=blue]{hyperref}

\DeclareMathOperator{\tr}{\text{tr}}

\renewcommand{\tfrac}{\genfrac{}{}{}1}
\newcommand{\ttfrac}{\genfrac{}{}{}2}

\newcommand{\Q}{\ensuremath{\mathcal{Q}}}
\newcommand{\Qb}{\ensuremath{\overline{\mathcal{Q}}}}

\newcommand{\bt}{\ensuremath{\overline{\theta}}}

\begin{document}

\thispagestyle{empty}

\begin{flushright}
CPHT-RR047.112016\\
DESY 16-206\\
\end{flushright}
\vskip .8 cm
\begin{center}
{\Large {\bf Effective field theory for magnetic compactifications}}\\[12pt]

\bigskip
\bigskip 
{
{\bf{Wilfried Buchmuller$^\dagger$}\footnote{E-mail: wilfried.buchmueller@desy.de}},
{\bf{Markus Dierigl$^\dagger$}\footnote{E-mail: markus.dierigl@desy.de}},  
{\bf{Emilian Dudas$^\ast$}\footnote{E-mail: emilian.dudas@cpht.polytechnique.fr}},
{\bf{Julian~Schweizer$^\dagger$}\footnote{E-mail: julian.schweizer@desy.de}}
\bigskip}\\[0pt]
\vspace{0.23cm}
{\it $^\dagger$ Deutsches Elektronen-Synchrotron DESY, 22607 Hamburg, Germany \\ \vspace{0.2cm}
$^\ast$ Centre de Physique Th\'eorique, \'Ecole Polytechnique, CNRS, Universit\'e Paris-Saclay, F-91128 Palaiseau, France}\\[20pt] 
\bigskip
\end{center}

\begin{abstract}
\noindent
Magnetic flux plays an important role in compactifications of field and string theories in two ways, it generates a multiplicity of chiral fermion zero modes and it can break supersymmetry. We derive the complete four-dimensional effective action for $\mathcal{N}=1$ supersymmetric Abelian and non-Abelian gauge theories in six dimensions compactified on a torus with flux. The effective action contains the tower of charged states and it accounts for the mass spectrum of bosonic and fermionic fields as well as their level-dependent interactions. This allows us to compute quantum corrections to the mass and couplings of Wilson lines. We find that the one-loop corrections vanish, contrary to the case without flux. This can be traced back to the spontaneous breaking of symmetries of the six-dimensional theory by the background gauge field, with the Wilson lines as Goldstone bosons.
\end{abstract}

\newpage 
\setcounter{page}{2}
\setcounter{footnote}{0}

{\renewcommand{\baselinestretch}{1.5}

\section{Introduction}
\label{sec:Introduction}

Magnetic flux plays a crucial role in the compactification of field theories and string theories\footnote{For a review and references see, for example \cite{Angelantonj:2002ct,Douglas:2006es,Blumenhagen:2006ci}.} in several ways. First of all, it leads to a multiplicity of fermion zero modes, which can be used to explain the number of quark-lepton generations \cite{Witten:1984dg}. Moreover, it is an important source of supersymmetry breaking \cite{Bachas:1995ik} and, together with a non-perturbative superpotential, it can stabilize the compact dimensions \cite{Braun:2006se} consistent with four-dimensional Minkowski or de Sitter vacua \cite{Buchmuller:2016dai,Buchmuller:2016bgt}.

In this paper we study the effect of flux on quantum corrections. We consider the simplest case, a six-dimensional (6d) gauge theory with $\mathcal{N}=1$ supersymmetry compactified to four dimensions (4d) on a torus $T^2$. Following techniques developed in \cite{Marcus:1983wb,ArkaniHamed:2001tb}, we start from the 6d Lagrangian written in terms of 4d chiral and vector superfields. Before we consider the constant magnetic flux background we derive a supersymmetric effective action for the Kaluza-Klein (KK) states of a 6d Abelian gauge theory and show how the KK excitations of the vector multiplet obtain their masses from a supersymmetric St\"uckelberg mechanism. In the flux background the covariant derivatives of the charged fields satisfy a harmonic oscillator algebra \cite{Bachas:1995ik,Braun:2006se,Alfaro:2006is,Abe:2012ya,Abe:2014noa}, which allows to encode their dynamics in the compact dimensions via ladder operators. Applying this harmonic oscillator analogy to the full superfields we derive a 4d supersymmetric effective action that incorporates the complete tower of charged states. Even though the explicit form of the field profiles in the flux background is known \cite{Bachas:1995ik,Cremades:2004wa} our analysis only uses their orthonormality. The 4d effective action contains the masses of all charged fields, which are reminiscent of Landau levels, as well as their interactions, see also \cite{Hamada:2012wj}. A similar treatment is carried out for non-Abelian flux where also components of the gauge field are charged and affected by the flux.

Internal magnetic fields were largely discussed in the string literature, see e.g. \cite{Angelantonj:2000hi, Dudas:2005jx}, starting with \cite{Bachas:1995ik}, followed by its T-dual interpretation of D-branes at angles \cite{Berkooz:1996km, Aldazabal:2000cn, Anastasopoulos:2011kr}. Global string theory models of this type with 4d supersymmetry or completely broken supersymmetry were constructed in \cite{Blumenhagen:2000wh}. In the case where 4d supersymmetry is broken by the internal magnetic flux, however, a NS-NS tadpole appears at the disk level, which signals a change of the ground state of the theory (see e.g.~\cite{Dudas:2004nd}). As a result, most quantum corrections at the string theory level cannot be reliably computed. On the other hand, the effective field theory action we construct in this paper is adapted to compute quantum corrections, as we will exemplify in the following. 

Our main interest concerns the effect of flux on the quantum corrections to massless scalar particles. Without flux it is well-known that the mass of Wilson lines, i.e.~the scalar zero modes associated with the higher-dimensional gauge field, are protected from quadratic divergences by the invariance under a discrete shift symmetry, a consequence of the higher-dimensional gauge invariance on the torus \cite{ArkaniHamed:2001nc,Antoniadis:2001cv}. For 6d gauge theories the one-loop effective potential has been explicitly computed in \cite{Antoniadis:2001cv}, and it has been shown that the squared mass of the Wilson line is proportional to the volume of the compact dimensions.

In the following we compute the one-loop correction to the Wilson line mass for an Abelian gauge theory in the magnetic flux background. As we shall see, due to the effect of the flux on the KK spectrum and couplings various cancellations occur, and the total one-loop mass vanishes. The same is true for the one-loop quartic coupling. This result can be understood by considering the 6d Lagrangian rather than the 4d Lagrangian: the Wilson lines are the Goldstone bosons of symmetries of the 6d Lagrangian, which are spontaneously broken by the background gauge field.

The paper is organized as follows. To introduce some formalism we first consider a supersymmetric Abelian 6d gauge theory without flux and derive the 4d Lagrangian of all KK modes in terms of chiral and vector superfields in Sec.~\ref{sec:Abeliannoflux}. The 4d Lagrangian in the case of flux is derived for an Abelian and a non-Abelian gauge theory in Sec.~\ref{sec:Abelianflux} and Sec.~\ref{sec:nonAbelianflux}, respectively. Sec.~\ref{sec:Radcorr} deals with the one-loop effective potential for the Wilson line of an Abelian gauge theory, and the mass and the quartic coupling of the Wilson line are computed in the case with flux. The role of Wilson lines as Goldstone bosons is discussed in Sec.~\ref{sec:goldstone}, and we conclude with an outlook in Sec.~\ref{sec:conclusion}.

\section{Abelian effective action without flux}
\label{sec:Abeliannoflux}

We consider a globally supersymmetric $U(1)$ gauge theory in six dimensions. Two of the dimensions are compactified on a square torus $T^2$ of area $L^2$. Following \cite{ArkaniHamed:2001tb}, we decompose the 6d vector multiplet into an $\mathcal{N}=1$ vector multiplet $V$ and a chiral multiplet $\phi$. The scalar part of $\phi$ contains the internal components of the vector field,
\begin{align}
\phi |_{\theta = \bt = 0} = \tfrac{1}{\sqrt{2}} (A_6 + i A_5) \,.
\label{scalarphi}
\end{align}
The six-dimensional gauge action can then be written in 4d superspace as \cite{ArkaniHamed:2001tb}
\begin{align}
S_6 = \int d^6 x \left\{ \frac{1}{4} \int d^2 \theta \, W^{\alpha}
  W_{\alpha} + \text{h.c.} + \int d^4 \theta \left( \partial V
    \overline{\partial} V + \phi \overline{\phi} + \sqrt{2} V
    \left(\overline{\partial} \phi + \partial \overline{\phi} \right)\right)\right\} \,,
\label{gaugeaction}
\end{align}
with $\partial = \partial_5 - i \partial_6$. Note that compared to \cite{ArkaniHamed:2001tb} we have performed an integration by parts in the last term. We further include a hypermultiplet of charge $q$ that decomposes into two 4d chiral multiplets of opposite charge, $Q$ and $\tilde{Q}$. The corresponding matter action can be written as
\begin{align}
S_6 = \int d^6 x \left\{ \int d^2 \theta \, \tilde{Q} (\partial + \sqrt{2} g q \phi ) Q + \text{h.c.} + \int d^4 \theta  \left( \overline{Q} e^{2 g q V} Q + \overline{\tilde{Q}} e^{-2 g q V} \tilde{Q} \right)\right\}\,.
\label{matteraction}
\end{align}

It is straightforward to compute the 4d effective action, keeping the full KK tower in the gauge sector as well as the matter sector. The superfields, which depend on all six coordinates, can be decomposed in terms of modes of fixed internal momenta,
\begin{equation}
\begin{split}
\phi (x_M, \theta, \bt) &= \sum_{n,m} \phi_{n,m} (x_{\mu}, \theta, \bt) \, \psi_{n,m} (x_m)\,, \\
V (x_M, \theta, \bt) &= \sum_{n,m} V_{n,m} (x_{\mu}, \theta, \bt) \, \psi_{n,m} (x_m)\,;
\label{vectorexpansion}
\end{split}
\end{equation}
here the index $M$ runs over all spacetime dimensions, whereas $\mu$ and $m$ only run over non-compact and compact dimensions, respectively. The $\psi_{n,m}$ are a complete set of mode functions that we choose as
\begin{align}
\psi_{n,m} (x_m) = \frac{1}{L} \exp \left[ \frac{2\pi i}{L} (n x_5 + m x_6 )\right] \,;
\end{align}
thay satisfy the orthonormality condition
\begin{align}
\int_{T^2} d^2 x \ \psi_{n,m} \overline{\psi}_{k,l} = \delta_{n,k} \delta_{m,l}\,.
\label{orthonorm1}
\end{align}
The reality of the vector field, $V=\overline{V}$, implies for the mode functions $\overline{V}_{n,m} = V_{-n,-m}$. Using the expansion \eqref{vectorexpansion} for vector and chiral superfields and integrating over the compact dimensions we obtain from Eq.~\eqref{gaugeaction} the equivalent 4d gauge action containing the full KK tower,
\begin{align}
\begin{split}
S_4 = \int d^4 x \sum_{n,m} &\bigg\{\int d^2 \theta \, \frac{1}{4}
  W^{\alpha}_{n,m} W_{\alpha,-n,-m} + \text{h.c.}  + \int d^4 \theta
  \Big( |M_{n,m}|^2 V_{n,m} \overline{V}_{n,m}
     \\
& \quad + \overline{\phi}_{n,m} \phi_{n,m} -
    \sqrt{2}\left(\overline{M}_{n,m} \overline{V}_{n,m} \phi_{n,m} +
      M_{n,m} V_{n,m} \overline{\phi}_{n,m} \right)\Big) \bigg\} \,,
\end{split}
\end{align}
where
\begin{align}
M_{n,m} = \frac{2\pi}{L}(m+in)\,.\label{Mnm}
\end{align}
The vector bosons of the KK tower acquire mass via the St\"uckelberg mechanism. At each KK level they absorb the imaginary part of of the complex field $\phi$ whereas the real part corresponds to the mass degenerate scalar that is needed to complete the massive vector multiplet. This can be made manifest by means of a shift of the vector field\footnote{For a discussion in component form see, for example \cite{Asaka:2001eh}.} ($M_{n,m} \neq 0$),
\begin{align}
V_{n,m} \rightarrow V_{n,m} +
\frac{1}{\sqrt{2}}\left(\frac{1}{M_{n,m}}\phi_{n,m} -
  \frac{1}{\overline{M}_{n,m}}\overline{\phi}_{-n,-m}\right)\,.
\end{align}
Performing the shift of $V_{n,m}$ and neglecting a total derivative, one obtains for the 4d gauge action 
\begin{align}
\begin{split}
S_4 = \int d^4 x \sum_{n,m} &\left\{ \int d^2 \theta \, \frac{1}{4}
  W^{\alpha}_{n,m} W_{\alpha,-n,-m} + \text{h.c.}  \right.\\
& \left. + \int d^4 \theta \left( |M_{n,m}|^2 V_{n,m} \overline{V}_{n,m} 
 + \overline{\varphi} \varphi \right)\right\}\,.
\end{split}
\end{align}
This is the standard $\mathcal{N}=1$ supersymmetric action for a massless vector multiplet together with a tower of massive KK vector multiplets. A massless chiral multiplet $\varphi \equiv \phi_{0,0}$ remains since the vector multiplet can only be shifted if $M_{n,m} \neq 0$.

In order to include the matter sector we have to evaluate integrals of three and four mode functions. This yields the couplings of the different KK levels and guarantees momentum conservation in the internal space. The relevant integrals are
\begin{equation}
\begin{split}
\int_{T^2} d^2 x \, \overline{\psi}_{n,m} \psi_{k,l} \psi_{r,s} &= \frac{1}{L} \, \delta_{n, k+r} \, \delta_{m, l+s} \,, \\
\int_{T^2} d^2 x \, \overline{\psi}_{n,m} \psi_{k,l} \psi_{r,s} \psi_{u,v} &= \frac{1}{L^2} \, \delta_{n, k+r+u} \,\delta_{m, l+s+v} \,.
\end{split}
\end{equation}
The complete effective 4d action, including gauge and matter KK towers, is given by
\begin{equation}
\begin{split}
S_4 =& \int d^4 x \sum_{n,m} \left\{ \int d^2 \theta \left( \frac{1}{4} W^{\alpha}_{n,m} W_{\alpha,-n,-m} + M_{n,m} \tilde{Q}_{n,m} Q_{n,m}\right) + \text{h.c.} \right. \\
&\left. + \int d^4 \theta \left( |M_{n,m}|^2\overline{V}_{n,m} V_{n,m}  +
  \overline{\phi}_{n,m} \phi_{n,m} + \overline{Q}_{n,m} Q_{n,m} +
  \overline{\tilde{Q}}_{n,m} \tilde{Q}_{n,m} \right)\right\}\\
&+ \int d^4x \sum_{n,m,k,l} \left\{ \int d^2 \theta \sqrt{2} q g \, \tilde{Q}_{n+k, m+l} \phi_{k,l} Q_{n,m}  + \text{h.c.}\right. \\
&\left. + \int d^4 \theta\ 2qg \left( \overline{Q}_{n+k, m+l} V_{k,l} Q_{n,m}  - \overline{\tilde{Q}}_{n,m} V_{k,l} \tilde{Q}_{n+k, m+l} \right) \right\}\\
&+ \int d^4 x \sum_{n,m,k,l,r,s} \left\{ \int d^4 \theta \, 2 q^2 g^2 \left( \overline{Q}_{n+k+r, m+l+s} V_{k,l} V_{r,s} Q_{n,m} \right. \right. \\
& \hspace{5.5cm} \left. \left. + \overline{\tilde{Q}}_{n,m} V_{k,l} V_{r,s} \tilde{Q}_{n+k+r, m+l+s} \right) \right\}\,.
\end{split}
\end{equation}
In addition to the gauge field $V_{n,m}$ and $\varphi$, it describes massless and massive matter fields that are formed from pairs of chiral multiplets $\tilde{Q}$ and $Q$ with Dirac mass terms $M_{n,m}$. At all KK levels the vector and matter fields are mass degenerate.

In the following sections we shall restrict the discussion to massless fields in the uncharged sector, which are denoted by $V_0$ and $\varphi$. The action then takes the
simplified form
\begin{equation}
\begin{split}
S_4^* =& \int d^4 x \left\{ \int d^2 \theta \left(
    \frac{1}{4} W^{\alpha}_0 W_{\alpha, 0} + \sum_{n,m} \left(M_{n,m} + \sqrt{2} qg 
      \varphi\right) \tilde{Q}_{n,m} Q_{n,m} \right) + \text{h.c.}\right. \\
 &\left. \hspace{1.8cm}+  \int d^4 \theta \left( \overline{\varphi} \varphi + \sum_{n,m} \left( \overline{Q}_{n,m} e^{2 q g V_0} Q_{n,m} + \overline{\tilde{Q}}_{n,m} e^{-2 q g V_0} \tilde{Q}_{n,m}\right)\right) \right\} \,.\label{S4-0}
\end{split}
\end{equation}
Note that the zero mode $\varphi$, the Wilson line of the gauge field, couples to matter like the mass terms.

\section{Abelian effective action with flux}
\label{sec:Abelianflux}

Now we turn the attention to the flux background in the internal dimensions. Since they are compact the flux is quantized. Moreover, the mass spectrum and field profiles of charged fields will be changed drastically and resemble that of Landau levels. Due to the magnetic field the charged fields will be localized in the extra dimensional space. A harmonic oscillator analogy, based on the work of \cite{Bachas:1995ik} and used in \cite{Braun:2006se,Alfaro:2006is,Abe:2014noa}, allows to explicitly construct the shape of the charged field profiles \cite{Bachas:1995ik,Cremades:2004wa}. In this way we obtain the four-dimensional effective action in terms of 4d superfields, restricted to the zero modes of the uncharged fields.

\subsection{Flux and the harmonic oscillator}
\label{subsec:fluxharmonic}

Before we derive the full supersymmetric effective action we want to elucidate the harmonic oscillator approach in a minimal example. For that reason we only consider the six-dimensional gauge field $A_M$ as a background for a charged scalar field $\Q$ of charge $q$. Consequently, the 6d action reads
\begin{align}
S_6 = \int d^6 x \, \left( - D_M \Qb D^M \Q \right) \,,
\label{6dbosonaction}
\end{align}
with the gauge covariant derivative acting as $D_M \Q = (\partial_M + i q g \, A_M) \Q$. The gauge field background accounts for a constant flux density $f$ in the internal dimensions, which in our choice of gauge reads\footnote{The calculations in the following sections are equally valid for other gauge choices.}
\begin{align}
A_5 = - \tfrac{1}{2} f x_6 \,, \quad A_6 = \tfrac{1}{2} f x_5 \,, \quad F_{56} = \partial_5 A_6 - \partial_6 A_5 = f \,.
\label{symmetricgauge}
\end{align}
As mentioned above, for the square torus of volume $L^2$ the flux is quantized. In the presence of particles with charge $q$ the flux density can take the values
\begin{align}
\frac{q g}{2\pi} \int_{T^2} F = \frac{q g}{2\pi} \int_{T^2} dx_5 dx_6 \, F_{56} = \frac{q g}{2\pi} L^2 f \in \mathbb{Z} 
\end{align}
Using a product space metric for $M_4 \times T^2$, and splitting the kinetic terms into 4d and 2d parts, the six-dimensional action \eqref{6dbosonaction} decomposes as
\begin{align}
S_6 = \int d^6x \left( - \eta^{\mu \nu} D_{\mu} \overline{\Q} D_{\nu} \Q - \overline{\Q}H_2 \Q \right) \,,
\end{align}
where after integration by parts in the internal coordinates we define the 2d Hamiltonian
\begin{align}
H_2 = -D_5^2 -D_6^2 = -\left( \partial_5 - \tfrac{i}{2} q g f x_6  \right)^2 - \left( \partial_6 + \tfrac{i}{2} q g f x_5 \right)^2 \,.
\end{align}
In analogy to the quantum harmonic oscillator with Hamiltonian $H = \tfrac{1}{2m} p^2 + \tfrac{1}{2} m \omega^2 x^2$ and the standard commutator relation $[x, p] = i \hbar$, we identify
\begin{align}
p = i D_6 \,, \quad x = i D_5 \,, \quad m = \tfrac{1}{2} \,, \quad \omega = 2 \,,
\end{align}
with the commutator relation
\begin{align}
[i D_5, i D_6 ] = -i q g f \,.
\end{align}
This leads to the further identification $\hbar = -q g f$ \cite{Braun:2006se}, since we choose $f$ to be negative for left-handed zero modes, c.f.~\cite{Buchmuller:2015eya}. One now defines the ladder operators
\begin{equation}
\begin{split}
a &= \sqrt{\frac{1}{-2qgf}} (i D_5 - D_6) \,, \\
a^{\dagger} &= \sqrt{\frac{1}{-2qgf}} (i D_5 + D_6) \,,
\label{ladderops}
\end{split}
\end{equation}
which satisfy the canonical commutator relation $[a, a^{\dagger}] = 1$. The internal Hamiltonian can be written in terms of the ladder operators as
\begin{align}
H_2 = - q g f \left( a^{\dagger} a + a a^{\dagger} \right) = - 2 q g f \left( a^{\dagger} a + \tfrac{1}{2} \right) \,.
\end{align}
Therefore, the energy eigenvalues of $H_2$ and thus the 4d Landau level masses show the typical spectrum of an harmonic oscillator. All levels are $|N|$-fold degenerate, with $N$ the number of flux quanta on the torus, in analogy to Landau levels. We denote the internal field profiles as $\psi_{n,j}$, see \cite{Cremades:2004wa}, where $n$ refers to the Landau level and $j$ accounts for the $|N|$-fold degeneracy. The field profiles corresponding to the lowest mass can then be constructed from the condition
\begin{align}
a \, \psi_{0,j} = 0 \,, \quad a^{\dagger} \, \overline{\psi}_{0,j} = 0 \,.
\end{align}
Applying the ladder operator we obtain the higher mode functions 
\begin{align}
\psi_{n,j} = \frac{1}{\sqrt{n!}} (a^{\dagger})^n \, \psi_{0,j}  \,, \quad \overline{\psi}_{n,j} = \frac{1}{\sqrt{n!}} (a)^n \, \overline{\psi}_{0,j} \,.
\end{align}
The explicit form of the lowest wave function was obtained in \cite{Bachas:1995ik,Cremades:2004wa}. In our consideration the specific form of the field profile is irrelevant and we will only need the orthonormality condition\footnote{Note that the charged wave functions in the flux background are not orthonormal with respect to the standard KK states discussed in Sec.~\ref{sec:Abeliannoflux}. Therefore, to discuss the interaction of the charged states with higher excitations of the uncharged sector one has to evaluate the overlaps explicitly, see e.g.~\cite{Hamada:2012wj}.}
\begin{align}
\int_{T^2} d^2 x \, \overline{\psi}_{\tilde{n},\tilde{\jmath}} \psi_{n,j} = \delta_{n, \tilde{n}} \delta_{j, \tilde{\jmath}} \,.
\label{orthonorm}
\end{align}

Instead of the KK decomposition in Sec.~\ref{sec:Abeliannoflux} we now decompose the charged fields with respect to the Landau levels,
\begin{equation}
\begin{split}
\Q (x_M) =& \sum_{n,j} \Q_{n,j} (x_{\mu}) \psi_{n,j} (x_m) = \sum_{n,j} \Q_{n,j} (x_{\mu}) \frac{1}{\sqrt{n!}} \left(a^{\dagger}\right)^n \psi_{0,j} (x_m) \,,\\
\Qb (x_M) =& \sum_{n,j} \Qb_{n,j} (x_{\mu}) \overline{\psi}_{n,j} (x_m) = \sum_{n,j} \Qb_{n,j} (x_{\mu}) \frac{1}{\sqrt{n!}} \left(a\right)^n\overline{\psi}_{0,j} (x_m) \,.
\end{split}
\end{equation}
The 6d action \eqref{6dbosonaction} then becomes
\begin{equation}
\begin{split}
S_6 =& \int d^4x \sum_{n,j,m,k} \left\{ - \eta^{\mu \nu} D_{\mu} \Qb_{n,j} D_{\nu} \Q_{m,k} \int_{T^2} d^2 x \, \overline{\psi}_{n,j} \psi_{m,k} \right. \\
& \left. \hspace{2.5cm} - \Qb_{n,j} \Q_{m,k} \int_{T^2} d^2 x \,(- 2 q g f) \overline{\psi}_{n,j}  \left(a^{\dagger} a + \tfrac{1}{2} \right) \psi_{m,k} \right\} \,.
\end{split}
\end{equation}
The four-dimensional effective action is derived by using the harmonic oscillator algebra and the orthonormality of the internal field profiles in the gauge field background,
\begin{align}
S_4 = \int d^4 x \sum_{n,j} \left( - D_{\mu} \Qb_{n,j} D^{\mu} \Q_{n,j} + (2qgf) \left( n + \tfrac{1}{2} \right) \Qb_{n,j}\Q_{n,j}\right) \,.
\end{align}
The masses for the 4d fields are given by
\begin{align}
m^2_{n,j} = - 2 q g f \left(n + \tfrac{1}{2} \right) = \frac{2 \pi |N|}{L^2} \left(2n +1 \right) \,,
\label{bosmass}
\end{align}
as discussed in \cite{Bachas:1995ik}. For fields with an internal helicity the mass formula is supplemented by a term $(-2qgf) \Sigma$, where $\Sigma$ is the internal helicity, see \cite{Bachas:1995ik}. This leads to the appearance of $|N|$ chiral fermion zero modes as predicted by the index theorem for the flux background ($\Sigma= \tfrac{1}{2}$) and a tachyonic mode in the presence of charged gauge fields with $\Sigma = 1$, as discussed in Sec.~\ref{sec:nonAbelianflux}.

\subsection{Supersymmetric effective action for Abelian flux}
\label{subsec:superAbelianflux}

The field profiles for charged fermions and bosons are identical because both arise as solutions to the gauge covariant Laplace equation on the torus. Therefore, instead of decomposing only the component fields with respect to the Landau levels we can decompose the superfield as a whole, similar to the procedure for the standard KK tower in Sec.~\ref{sec:Abeliannoflux}. As mentioned above, the six-dimensional hypermultiplet can be written in terms of two chiral multiplets with opposite charge,
\begin{equation}
\begin{split}
Q (x_M, \theta, \bt) &= \sum_{n,j} Q_{n,j}(x_{\mu}, \theta, \bt) \, \psi_{n,j}(x_m) \,, \\
\tilde{Q} (x_M, \theta, \bt) &= \sum_{n,j} \tilde{Q}_{n,j}(x_{\mu}, \theta, \bt) \, \overline{\psi}_{n,j}(x_m) \,.
\label{fluxmodeexp}
\end{split}
\end{equation}
Furthermore, the index theorem guarantees $|N|$ fermion zero modes. In our convention, c.f.~\cite{Buchmuller:2015eya}, we choose $f$ to be negative which corresponds to zero modes contained in the $\tilde{Q}$ multiplet.

The uncharged 6d vector multiplet has the usual KK expansion on the torus, see Sec.~\ref{sec:Abeliannoflux}. Here, we concentrate on its background value and the zero mode, which are encoded in $V_0$ and $\phi_0$. The scalar component of $\phi_0$ contains the internal component of the gauge field and therefore encodes the magnetic flux on the torus. We split this contribution into the background gauge field generating the flux and perturbations $\varphi$, which are constant with respect to the internal dimensions. Hence, $\varphi$ corresponds to the continuous Wilson lines on the torus, $ \varphi = \tfrac{1}{\sqrt{2}} (a_6 + i a_5)$. In the symmetric gauge \eqref{symmetricgauge} the scalar component reads
\begin{align}
\phi_0|_{\theta= \overline{\theta} = 0} = \frac{f}{2 \sqrt{2}} \left( x_5 - i  x_6 \right) + \varphi \,.
\end{align}
The coupling of the hypermultiplet to the internal components of the 6d gauge field can then be written in $\mathcal{N}=1$ notation as in Eq.~\eqref{matteraction}. Plugging in the expressions for the ladder operators \eqref{ladderops} and the mode expansion \eqref{fluxmodeexp}, we obtain
\begin{align}
S_6 \supset & \int d^6 x \int d^2 \theta \, \tilde{Q} (\partial + \sqrt{2} q g \phi_0) Q + \text{h.c.}  \nonumber \\
= & \int{d^6 x \int d^2 \theta \left(  - i \sqrt{-2qgf} \, \tilde{Q} a^{\dagger} Q + \sqrt{2} q g \, \tilde{Q} \, \varphi \, Q \right) + \text{h.c.} } \\
= &  \int{d^4 x \int d^2 \theta \sum_{n,\tilde{n},j,\tilde{\jmath}} \tilde{Q}_{\tilde{n},\tilde{\jmath}} \, Q_{n,j} \int_{T^2} d^2 x \left( \overline{\psi}_{\tilde{n},\tilde{\jmath}} \, (-i \sqrt{-2qgf} \, a^{\dagger} + \sqrt{2} q g\varphi)\psi_{n,j} \right) + \text{h.c.} }  \nonumber
\end{align}
After using the orthonormality condition \eqref{orthonorm} of the states, we find the contribution to the 4d effective action after integration over the torus,
\begin{align}
S_{4}^* & \supset \int d^4 x \int d^2 \theta \, W + \text{h.c.}  \label{superpot}\\
  & = \int d^4 x \int d^2 \theta \sum_{n,j} \left( -i \sqrt{-2qgf (n+1)} \tilde{Q}_{n+1,j} \, Q_{n,j} + \sqrt{2} qg \, \tilde{Q}_{n,j} \, \varphi \, Q_{n,j} \right) + \text{h.c.}  \nonumber
\end{align}
The superpotential contains a mass term for the charged superfields and an interaction term which couples them to the internal components of the gauge field, i.e.\ the Wilson lines. The kinetic terms of the charged fields can be treated similarly, which yield
\begin{equation}
\begin{split}
S_4^* &\supset \int d^4 x \int d^4 \theta \int_{T^2} d^2x\left( \overline{Q} e^{2qgV_0} Q + \overline{\tilde{Q}}  e^{-2qgV_0} \tilde{Q} \right) \\
&= \int d^4 x \int d^4 \theta \sum_{n,j} \left( \overline{Q}_{n,j} e^{2qgV_0} Q_{n,j} + \overline{\tilde{Q}}_{n,j} e^{-2qgV_0} \tilde{Q}_{n,j} \right) \,.
\end{split}
\end{equation}
Finally, the 4d zero modes of the gauge field are included, leading to the same effective action as in Sec.~\ref{sec:Abeliannoflux},
\begin{equation}
\begin{split}
S_4^* \supset & \int d^4 x \int_{T^2 } d^2x \left( \int d^2 \theta \, \frac{1}{4} W^{\alpha} W_{\alpha} + \text{h.c.}\right) \\
= &  \int d^4 x \left( \int d^2 \theta \, \frac{1}{4} W_0^{\alpha} W_{\alpha, 0} + \text{h.c.} \right) \,.
\end{split}
\end{equation} 
The last contribution we have to add leads to a kinetic term for the complex Wilson line $\varphi$ as well as a Fayet-Iliopoulos (FI) term\footnote{Here, we use $\overline{\partial} \phi = \partial \overline{\phi}= f/\sqrt{2}$ in the flux background, since $\partial \overline{\varphi} = 0 = \overline{\partial} \varphi$, and $\partial V = 0 = \overline{\partial} V$.}
\begin{equation}
\begin{split}
S_4^* &\supset \int d^4 x \int_{T^2} d^2 x \int d^4 \theta \left( \partial V_0 \overline{\partial} V_0 + \overline{\phi}_0 \phi_0 + \sqrt{2} V_0 \overline{\partial} \phi_0 + \sqrt{2} V_0 \partial \overline{\phi}_0 \right) \\
&= \int d^4 x \int d^4 \theta \, \left( \overline{\varphi} \varphi  + 2f V_0 \right)\,.
\end{split}
\end{equation}
Note again that compared to \cite{ArkaniHamed:2001tb} our action differs by an integration by parts. This is important  since the boundary terms do not vanish in the flux background. In summary, the 4d effective action with the complete tower of charged states and a restriction to the zero modes in the uncharged sector reads
\begin{align}
S_4^* = \int d^4 x & \left[ \int d^4 \theta \, \left( \overline{\varphi} \varphi + \sum_{n,j} (\overline{Q}_{n,j} e^{2gqV_0} Q_{n,j} + \overline{\tilde{Q}}_{n,j} e^{-2qgV_0} \tilde{Q}_{n,j}) + 2f V_0 \right) \right. \nonumber \\
  + &\int d^2 \theta \, \left( \frac{1}{4} W_0^{\alpha} W_{\alpha,0} \right. \\
& + \left. \left. \sum_{n,j} \left(-i \sqrt{-2qgf (n+1)} \tilde{Q}_{n+1,j} Q_{n,j} + \sqrt{2} qg \tilde{Q}_{n,j} \, \varphi \, Q_{n,j} \right) \right) + \text{h.c.} \right] . \nonumber
\label{effectiveaction}
\end{align}

In order to obtain the mass spectrum of the charged fields and their interactions with the uncharged field $\varphi$ one has to integrate out the auxiliary fields. The bosonic mass terms receive contributions from $F$- and $D$-terms, whereas only the $F$-terms enter for the fermion masses. The couplings of the auxiliary field $D$ are given by
\begin{align}
\mathcal{L}_D = f D + |\Q_{n,j}|^2 q g D - |\tilde{\Q}_{n,j}|^2 q g D + \frac{1}{2} D^2 \,,
\end{align}
yielding
\begin{align}
D = - f - q g \sum_{n,j} \left( |\Q_{n,j}|^2 - |\tilde{\Q}_{n,j}|^2 \right) \,.
\end{align}
Similarly, the $F$-terms appear in the component action as 
\begin{equation}
\begin{split}
\mathcal{L}_F = & |F_{\varphi}|^2 + \sum_{n,j} \left( |F_{n,j}|^2 + |\tilde{F}_{n,j}|^2 \right) \\
 + & \sum_{n,j} \left( -i \sqrt{-2qgf (n+1)} \left( \tilde{F}_{n+1,j} \Q_{n,j} + \tilde{\Q}_{n+1,j} F_{n,j} \right) \right. \\
 & +\left. \sqrt{2} q g \left( \tilde{F}_{n,j} \, \varphi \, \Q_{n,j} + \tilde{\Q}_{n,j} F_{\varphi} \Q_{n,j} + \tilde{\Q}_{n,j} \, \varphi \, F_{n,j} \right) \right) + \text{h.c.} \,,
\end{split}
\end{equation}
leading to
\begin{equation}
\begin{split}
F_{n,j} &= - i \sqrt{-2 q g f (n+1)} \, \overline{\tilde{\Q}}_{n+1,j} - \sqrt{2} q g \, \overline{\tilde{\Q}}_{n,j} \,\overline{\varphi} \,, \\
\tilde{F}_{n+1,j} &= - i \sqrt{-2qgf (n+1)} \, \overline{\Q}_{n,j} - \sqrt{2} q g \, \overline{\Q}_{n+1,j} \, \overline{\varphi}\,, \\
F_{\varphi} &= - \sqrt{2} q g \sum_{n,j} \overline{\tilde{\Q}}_{n,j} \overline{\Q}_{n,j} \,.
\end{split}
\end{equation}
Plugging the $F$- and $D$-terms back into the component Lagrangian we find the bosonic mass terms
\begin{align}
\mathcal{L}_M^b = - \sum_{n,j} \left[ -2qgf (n+1) \left( |\tilde{\Q}_{n+1,j}|^2 + |\Q_{n,j}|^2 \right) + q g f |\Q_{n,j}|^2 - q g f|\tilde{\Q}_{n,j}|^2 \right] \,.
\end{align}
Therefore, the two scalars of different charge have the same tower of massive states due to a charge dependent shift induced by the $D$-term, leading to the bosonic masses evaluated in Eq.~\eqref{bosmass},
\begin{equation}
\begin{split}
m_{\tilde{\Q}_{n,j}}^2 &= -2qgfn -qgf = -qgf (2n+1) = \frac{2 \pi |N|}{L^2} (2n+1) \,, \\
m_{\Q_{n,j}}^2 &= -2qgf (n+1) + qgf = \frac{2 \pi |N|}{L^2} (2n+1) \,.
\end{split}
\end{equation}
The fermionic mass terms can be directly read off the superpotential \eqref{superpot},
\begin{align}
\mathcal{L}_M^f = \sum_{n,j} \left[ -i \sqrt{2qgf (n+1)} \tilde{\chi}_{n+1,j} \chi_{n,j} + \text{h.c.} \right] \,.
\end{align}
We find the $|N|$ chiral zero modes $\tilde{\chi}_{0,j}$ predicted by the index theorem. The rest of the chiral fermions pair up to form massive Dirac fields,
\begin{align}
\Psi_{n,j} = \begin{pmatrix} \tilde{\chi}_{n+1, j} \\ \chi_{n,j} \end{pmatrix} \,,
\end{align}
with masses
\begin{align}
m_{\Psi_{n,j}}^2 = -2qgf (n+1) = \frac{4 \pi |N|}{L^2} (n+1) \,.
\end{align}
Including the interactions among fermions and bosons we arrive at the full component Lagrangian 
\begin{align}
\mathcal{L}_{eff} = \mathcal{L}_{kin} + \mathcal{L}_M + \mathcal{L}_{int} - \tfrac{1}{2} f^2  \,,
\end{align}
with the bilinear kinetic and mass terms
\begin{equation}
\begin{split}
\mathcal{L}_{kin} = & - \partial_{\mu} \overline{\varphi} \partial^{\mu} \varphi - \frac{1}{4} F_{\mu \nu} F^{\mu \nu} - \sum_{n,j} \left( D_{\mu} \overline{\Q}_{n,j} D^{\mu} \Q_{n,j} + D_{\mu}\overline{\tilde{\Q}}_{n,j} D^{\mu} \tilde{\Q}_{n,j} \right) \\
& - i \left( \lambda_1 \sigma^{\mu} \partial_{\mu} \overline{\lambda}_1 + \lambda_2 \sigma^{\mu} \partial_{\mu} \overline{\lambda}_2 \right) - i \sum_{n,j} \left( \chi_{n,j} \sigma^{\mu} D_{\mu} \overline{\chi}_{n,j} + \tilde{\chi}_{n,j} \sigma^{\mu} D_{\mu} \overline{\tilde{\chi}}_{n,j} \right) \,, 
\end{split}
\end{equation}
\begin{equation}
\begin{split}
\mathcal{L}_{M} = & - \sum_{n,j} (-2 q g f) \left( n + \tfrac{1}{2} \right) \left( |\Q_{n,j}|^2 + |\tilde{\Q}_{n,j}|^2 \right) \\
& + i \sum_{n,j} \sqrt{-2qgf (n+1)} \tilde{\chi}_{n+1,j} \chi_{n,j} + \text{h.c.} \,, 
\label{massL}
\end{split}
\end{equation}
and the cubic and quartic interaction terms
\begin{equation}
\begin{split}
\mathcal{L}_{int} = & - \frac{g^2 q^2}{2} \left[ \sum_{n,j} \left( |\Q_{n,j}|^2 - |\tilde{\Q}_{n,j}|^2\right) \right]^2 - 2 q^2 g^2 \left(\sum_{n,j} \overline{\tilde{\Q}}_{n,j} \overline{\Q}_{n,j} \right) \left(\sum_{m,k} \tilde{\Q}_{m,k} \Q_{m,k} \right) \\
& - i \sqrt{2} q g \sum_{n,j} \sqrt{-2q gf (n+1)} \left( \overline{\tilde{\Q}}_{n+1,j} \, \varphi \, \tilde{\Q}_{n,j} - \overline{\Q}_{n,j} \, \varphi \, \Q_{n+1,j} \right) + \text{h.c.} \\
& - 2 q^2 g^2 \sum_{n,j} |\varphi|^2 \left( |\Q_{n,j}|^2 + |\tilde{\Q}_{n,j}|^2 \right) \\
& + \sqrt{2} q g  \sum_{n,j} \left( i \overline{\Q}_{n,j} \lambda_1 \chi_{n,j} - i \overline{\tilde{\Q}}_{n,j} \lambda_1 \tilde{\chi}_{n,j} - \Q_{n,j} \lambda_2 \tilde{\chi}_{n,j} - \tilde{\Q}_{n,j} \lambda_2 \chi_{n,j} \right) + \text{h.c.} \\
& - \sqrt{2} q g \sum_{n,j} \varphi \tilde{\chi}_{n,j} \chi_{n,j} + \text{h.c.} 
\label{effectivecomponent}
\end{split}
\end{equation}
Note that the fermions $\tilde{\chi}_{0,j}$, $j \in \{ 1, \dots, |N| \}$, are the only charged massless fields.

\section{Non-Abelian flux background}
\label{sec:nonAbelianflux}

In the case of non-Abelian flux we proceed very similar to the Abelian case above. In the following we will consider a six-dimensional super Yang-Mills theory with gauge group $SU(2)$. The generalization to higher rank gauge groups and the inclusion of charged matter fields is straightforward. Again, the starting point is the 6d action expressed in 4d superfields as given in \cite{ArkaniHamed:2001tb}. Moreover, we will always work in the Wess-Zumino (WZ) gauge. The fields of the 6d non-Abelian theory are contained in a vector multiplet $V$ and a chiral multiplet $\phi$ that both transform in the adjoint representation,
\begin{align}
S_6 =\int d^6 x &\left\{ \frac{1}{2} \int d^2 \theta \tr \left( W^{\alpha} W_{\alpha} \right) + \text{h.c.} \right. \\
& + \left. \int d^4 \theta \frac{2}{g^2} \tr \left( \left(\sqrt{2} \, \overline{\partial} + g \overline{\phi} \right) e^{-gV} \left(-\sqrt{2} \, \partial + g \phi \right) e^{gV} + \overline{\partial} e^{-gV} \partial e^{gV}\right) \right\} \nonumber \,,
\end{align}
with the trace convention $\tr \left( T_a T_b \right) = \tfrac{1}{2} \delta_{ab}$. Expanding the exponentials, integrating some of the terms by part, and bearing in mind that $V^3 = 0$ in the WZ gauge, this action can be written as
\begin{align}
S_6 = \int d^6 x & \left\{ \frac{1}{2} \int d^2 \theta \tr \left( W^{\alpha} W_{\alpha }\right) + \text{h.c.} \right. \nonumber \\
+ & \int d^4 \theta \bigg[ 2 \tr \left( \overline{\phi} \phi + \sqrt{2} \left(\partial \overline{\phi} + \overline{\partial} \phi \right) V \right) \\
& + \left.  2 \tr \left( g \left[ \, \overline{\phi}, \phi \right] V + \left(\overline{\partial} V - \frac{g}{\sqrt{2}} \left[ V, \overline{\phi} \, \right]\right) \left(\partial V + \frac{g}{\sqrt{2}} \left[ V, \phi \right] \right) \right) \bigg] \right\} \nonumber \,.
\end{align}
In order to evaluate the group structure we define a new basis of generators $\{T_3, T_+, T_-\}$, where $T_i$ are the properly normalized Pauli matrices and $T_{\pm} = T_1 \pm i T_2$. In this basis one has
\begin{equation}
\begin{split}
& \tr \left( T_{\pm}^2 \right) = 0 \,, \quad \tr \left( T_+ T_- \right) = 1 \,, \quad \tr \left( T_{\pm} T_3 \right) = 0 \,, \quad \tr \left(T_3^2\right) = \tfrac{1}{2} \,, \\
& \left[ T_+, T_- \right] = 2 T_3 \,, \quad \left[ T_3, T_{\pm} \right] = \pm T_{\pm} \label{commutator} \,.
\end{split}
\end{equation}
The chiral field $\phi$ decomposes as
\begin{equation}
\begin{split}
\phi &= \phi_3 T_3 + \phi_+ \tfrac{1}{\sqrt{2}} T_- + \phi_- \tfrac{1}{\sqrt{2}} T_+ \,, \\
\overline{\phi} &= \overline{\phi}_3 T_3 + \overline{\phi}_+ \tfrac{1}{\sqrt{2}} T_+ + \overline{\phi}_- \tfrac{1}{\sqrt{2}} T_- \,,
\label{nonabeliandecomposition}
\end{split}
\end{equation}
and analogously the vector field, whose reality condition leads to $\overline{V}_3 = V_3$ and $\overline{V}_{\pm} = V_{\mp}$. The flux will be encoded as non-trivial background for the field $\phi_3$ corresponding to the Cartan generator $T_3$, similar to the Abelian case \eqref{symmetricgauge},
\begin{align}
\phi_3 = \frac{f}{2 \sqrt{2}} (x_5 - i x_6) + \varphi_3 \,.
\label{nonabeliansymmetricgauge}
\end{align}
The commutator identities \eqref{commutator} show that the components $\phi_+,\,V_+$ and $\phi_-,\, V_-$ have positive and negative charge with respect to $V_3$, respectively. Consequently, their internal field profiles will be the same as the charged wave functions in the Abelian case. In our normalization convention the charge of the chiral fields $\phi_{\pm}$ is $q = \pm \tfrac{1}{2}$. The fields $V_3$ and $\phi_3$ are uncharged and we will only take their 4d zero modes into account\footnote{In the following the restriction to the zero mode for $V_3$ is understood and we do not indicate this with a subscript $0$. The zero mode of $\phi_3$ is denoted by $\varphi_3$ similar to the previous sections.}. Since the wave functions are identical to the Abelian framework we can also adopt the harmonic oscillator analogy and define the ladder operators in terms of the background gauge field, c.f.~\eqref{ladderops}
\begin{align}
a^{\dagger} = \frac{i}{\sqrt{-gf}} \left( \partial + \frac{g}{\sqrt{2}} (\phi_3 - \varphi_3) \right) \,, \quad a = \frac{i}{\sqrt{-gf}} \left( \overline{\partial} - \frac{g}{\sqrt{2}} (\overline{\phi}_3 - \overline{\varphi}_3) \right) \,.
\end{align}
The full six-dimensional action can then be expressed in the basis \eqref{nonabeliandecomposition}. After integration by parts one obtains
\begin{equation}
\begin{split}
S_6 = \int d^6 x &\left\{ \int d^2 \theta \left( \frac{1}{4} W^{\alpha}_3 W_{\alpha,3} + \frac{1}{2} W^{\alpha}_+ W_{\alpha, -}\right) + \text{h.c.}\right. \\
& + \int d^4 \theta \bigg (\overline{\varphi}_3 \varphi_3 + \overline{\phi}_+ e^{g V_3} \phi_+ + \overline{\phi}_- e^{-g V_3} \phi_-  + 2 f V_3 \\
& \hspace{1.8cm} + V_-  \left(i \sqrt{-gf} a^{\dagger} - \tfrac{g}{\sqrt{2}} \varphi_3  \right)\left( -i \sqrt{-gf} a - \tfrac{g}{\sqrt{2}} \overline{\varphi}_3 \right) V_+ \\
& \hspace{1.8cm} + V_-  \left(i \sqrt{-gf} a + \tfrac{g}{\sqrt{2}} \overline{\varphi}_3  \right)\left( -i \sqrt{-gf} a^{\dagger} + \tfrac{g}{\sqrt{2}} \varphi_3 \right) V_+ \\
& \hspace{1.8cm} - \sqrt{2} V_- \left( 1 - \tfrac{g}{\sqrt{2}}  V_3 \right) \left( i \sqrt{-gf} a^{\dagger} - \tfrac{g}{\sqrt{2}} \varphi_3 \right) \overline{\phi}_- \\
& \hspace{1.8cm} - \sqrt{2} \phi_- \left( 1 - \tfrac{g}{\sqrt{2}}  V_3 \right) \left( -i \sqrt{-gf} a - \tfrac{g}{\sqrt{2}} \overline{\varphi}_3 \right) V_+ \\
& \hspace{1.8cm} - \sqrt{2} \overline{\phi}_+ \left( 1 + \tfrac{g}{\sqrt{2}}  V_3 \right) \left(- i \sqrt{-gf} a^{\dagger} + \tfrac{g}{\sqrt{2}} \varphi_3 \right) V_+ \\
& \hspace{1.8cm} - \sqrt{2} V_- \left( 1 + \tfrac{g}{\sqrt{2}}  V_3 \right) \left( i \sqrt{-gf} a + \tfrac{g}{\sqrt{2}} \overline{\varphi}_3 \right) \phi_+ \\
& \hspace{1.8cm} + \tfrac{g^2}{2} \left( V_+ \phi_- - V_- \phi_+ \right) \left( V_- \overline{\phi}_- - V_+ \overline{\phi}_+ \right) \bigg) \bigg\} \,.
\end{split}
\end{equation}
We clearly identify the kinetic term for $\varphi_3$ as well as the gauge covariant kinetic terms for the charged chiral multiplets $\phi_{\pm}$ of charge $\pm \tfrac{1}{2}$. Also the FI-term for the vector multiplet aligned with the flux $V_3$ is present, as in the Abelian case. The remaining contributions will lead to interaction and mass terms connecting different charged states. Except the last term, that contains four charged fields, we can derive the 4d effective action along the lines of Sec.~\ref{sec:Abelianflux}, where $\phi_{\pm}$ now correspond to the charged chiral multiplets $Q$ and $\tilde{Q}$. The final result is
\begin{align}
S_4^* = \int d^4 x & \left\{ \int d^2 \theta \left( \frac{1}{4} W^{\alpha}_3 W_{\alpha,3} +  \frac{1}{2} \sum_{n,j}  W^{\alpha}_{+,n,j} W_{\alpha,-,n,j}\right) + \text{h.c.} \right. \nonumber \\
& + \int d^4 \theta \left[ \overline{\varphi}_3 \varphi_3  + 2 f V_3 + \sum_{n,j} \left( \overline{\phi}_{+,n,j} e^{g V_3} \phi_{+,n,j} + \overline{\phi}_{-,n,j} e^{-g V_3} \phi_{-,n,j}\right)\right. \nonumber \\
& \hspace{0.4cm} + \sum_{n,j} \left( (2n+1)(-gf) V_{-,n,j}V_{+,n,j} + i \sqrt{2n (-gf)} g \varphi_3 V_{-,n-1,j} V_{+,n,j}\right. \nonumber \\
& \hspace{1.0cm} \left. -i \sqrt{2 (n+1) (-gf)} g \overline{\varphi}_3 V_{-,n+1,j} V_{+,n,j} + g^2 \overline{\varphi}_3 \varphi_3  V_{-,n,j} V_{+,n,j} \right) \nonumber \\
& \hspace{0.4cm} + \sum_{n,j} \left( \left( 1 - \tfrac{g}{\sqrt{2}} V_3 \right) \left( - i \sqrt{2(n+1)(-gf)} V_{-,n+1,j} \overline{\phi}_{-,n,j} \right.\right. \label{nonabelianaction}\\
& \hspace{1.0cm} \left. + i \sqrt{2 n (-gf)}  \phi_{-,n-1,j} V_{+,n,j} + g \varphi_3 V_{-,n,j} \overline{\phi}_{-,n,j} + g \overline{\varphi}_3 \phi_{-,n,j} V_{+,n,j} \right) \nonumber \\
& \hspace{1.8cm} + \left(1 + \tfrac{g}{\sqrt{2}} V_3\right) \left( i \sqrt{2(n+1)(-gf)} \overline{\phi}_{+,n+1,j} V_{+,n,j} \right. \nonumber \\
& \hspace{1.0cm} \left.\left. - i \sqrt{2 n (-gf)} V_{-,n-1,j} \phi_{+,n,j} - g \varphi_3 \overline{\phi}_{+,n,j} V_{+,n,j} - g \overline{\varphi}_3 V_{-,n,j} \phi_{+,n,j} \right)\right) \nonumber \\
& \hspace{0.4cm} + \left. \left. \sum_I \frac{g^2}{2} C_I \left(V_{+,n,j} \phi_{-,\tilde{n},\tilde{\jmath}} - V_{-,\tilde{n},\tilde{\jmath}} \phi_{+,,n,j}\right) \left( V_{-, \tilde{m},\tilde{l}} \overline{\phi}_{-,m,l} - V_{+,m,l} \overline{\phi}_{+, \tilde{m}, \tilde{l}}\right) \right] \right\} , \nonumber
\end{align}
with $I = \{n,j,\tilde{n}, \tilde{\jmath}, m, l, \tilde{m}, \tilde{l}\}$ and 
\begin{align}
C_I = \int_{T^2} d^2x \left( \psi_{n,j} \overline{\psi}_{\tilde{n}, \tilde{\jmath}} \psi_{m,l} \overline{\psi}_{\tilde{m}, \tilde{l}}\right)\,.
\end{align}
Integrating out the auxiliary fields in \eqref{nonabelianaction} we can work out the masses of the charged fields. The charged vector boson masses in 4d can be evaluated using $\theta \sigma^{\mu} \bt \theta \sigma^{\nu} \bt = - \tfrac{1}{2} \theta \theta \bt \bt \eta^{\mu \nu}$. They are given by
\begin{align}
m_{A_{\pm},n,j}^2 = \tfrac{1}{2} (-gf) (2 n + 1) = \frac{2\pi |N|}{L^2} (2n + 1)\,.
\end{align}
Hence, the charged vector fields have to absorb part of the charged scalar fields via the St\"uckelberg mechanism, c.f.~Sec.~\ref{sec:Abeliannoflux}. The necessary couplings of the charged gauge fields to the derivative of the scalars can be extracted from the action \eqref{nonabelianaction}
\begin{equation}
\begin{split}
S_4^* \supset & \int d^4 x \int d^4 \theta \sum_{n,j} \left[ - i \sqrt{-2 gf} \left(  \sqrt{n} \, \overline{\phi}_{-,n-1,j} + \sqrt{n+1} \, \phi_{+, n+1, j} \right) V_{-,n,j} \right. \\
& \hspace{3.1cm} + \left. i \sqrt{-2gf} \left( \sqrt{n}\, \phi_{-, n-1,j} + \sqrt{n+1} \, \overline{\phi}_{+,n+1 ,j} \right) V_{+,n,j}\right] \\
\supset & \int d^4 x \sum_{n,j} \sqrt{\frac{-gf}{2}} \left[ \left(- \sqrt{n} \, \partial_{\mu} \overline{\phi}_{-, n-1, j} + \sqrt{n+1} \, \partial_{\mu} \phi_{+, n+1, j} \right) A^{\mu}_{-,n,j} \right. \\
& \hspace{2.9cm} + \left. \left( - \sqrt{n} \, \partial_{\mu} \phi_{-,n-1,j} + \sqrt{n+1} \, \partial_{\mu} \overline{\phi}_{+, n+1, j}\right) A^{\mu}_{+,n,j}\right] \,.
\end{split}
\end{equation}
This identifies the eaten complex Goldstone mode\footnote{Here, we denote the scalar component of the superfields $\phi_{\pm}$ with the same letter as the superfield.}
\begin{align}
\Phi_{n,j} =& -\sqrt{\frac{n+1}{2n+3}} \overline{\phi}_{-,n,j} + \sqrt{\frac{n+2}{2n+3}} \phi_{+,n+2,j} \,,
\label{fluxgoldstone}
\end{align}
for the charged vector bosons $A_{\pm,n+1,j}^{\mu}$. The modes $A_{\pm,0,j}^{\mu}$ eat the complex bosons $\phi_{+,1,j}$.

To determine the mass spectrum for the remaining two real charged degrees of freedom we need to evaluate the $D$-terms. The solutions of the $D$-term equations read
\begin{align}
D_3 = - f - \frac{g}{2} \sum_{n,j} \left( \left| \phi_{+,n,j} \right|^2 - \left| \phi_{-,n,j}\right|^2 \right)  \,,
\end{align}
\begin{equation}
\begin{split}
D_{+,n,j} =& \, i \sqrt{\frac{-gf}{2}} \sqrt{2n+1} \left( \sqrt{\frac{n}{2n+1}} \overline{\phi}_{-,n-1,j} + \sqrt{\frac{n+1}{2n+1}} \phi_{+,n+1,j}\right) \,, \\
D_{-,n,j} =& - i \sqrt{\frac{-gf}{2}} \sqrt{2n+1} \left( \sqrt{\frac{n}{2n+1}} \phi_{-,n-1,j} + \sqrt{\frac{n+1}{2n+1}} \overline{\phi}_{+,n+1,j}\right) \,.
\end{split}
\end{equation}
Substituting the $D$-terms into the component action we can extract the quadratic part of the scalar Lagrangian and identify the mass terms
\begin{equation}
\begin{split}
\mathcal{L}_{M} \supset& - \frac{gf}{2} \left\{ \left| \phi_{+,0,j}\right|^2 \right.  \\
& - \left. \sum_{n,j} \begin{pmatrix} \overline{\phi}_{-,n,j}, & \overline{\phi}_{+,n+2,j} \end{pmatrix} \begin{pmatrix} n+2 & \sqrt{(n+1)(n+2)} \\ \sqrt{(n+1)(n+2)} & n+1 \end{pmatrix} \begin{pmatrix} \phi_{-,n,j} \\ \phi_{+,n+2,j}\end{pmatrix} \right\}
\label{nonabelianmasses}
\end{split}
\end{equation}
Since we chose $f < 0$ we see that there are $|N|$ tachyonic modes $\phi_{+,0,j}$ that will acquire vacuum expectation values in the true vacuum (see the comments at the end of this section). This corresponds to the helicity dependent mass shift one expects, as pointed out in Sec.~\ref{subsec:fluxharmonic}. The states $\phi_{+,1,j}$ have vanishing mass as should be the case for a St\"uckelberg field for the first level of massive gauge bosons. The rest of the tower has masses corresponding to the eigenvalues of the matrix in \eqref{nonabelianmasses}. Clearly, the determinant of the mass matrix vanishes, which indicates the massless Goldstone modes \eqref{fluxgoldstone}. The remaining complex scalar degree of freedom corresponds to the linear combination orthogonal to \eqref{fluxgoldstone},
\begin{align}
\tilde{\Phi}_{n,j} = \sqrt{\frac{n+2}{2n+3}} \overline{\phi}_{+,n+2,j} + \sqrt{\frac{n+1}{2n+3}} \phi_{-,n,j} \,,
\end{align}
with mass eigenvalues
\begin{align}
m^2_{\tilde{\Phi}_{n,j}} = \tfrac{1}{2}(-gf) (2n+3) = \frac{2\pi |N|}{L^2} (2n+3)\,.
\end{align}
Note that the physical mass spectrum differs from the one given in \cite{Bachas:1995ik}. The states with mass squared $2\pi|N| / L^2$ are absorbed by charged vector bosons. The fermion mass terms are
\begin{align}
\mathcal{L}_M \supset - \sum_{n,j} \sqrt{(n+1) (-gf)} \left( \lambda_{+,n+1,j} \tilde{\lambda}_{-,n,j}  - \lambda_{-,n,j} \tilde{\lambda}_{+,n+1,j} \right) + \text{h.c.} \,,
\end{align}
where $\lambda$ and $\tilde{\lambda}$ denote the gauginos contained in the vector multiplet and chiral multiplets, respectively. We find $2|N|$ fermion zero modes $\lambda_{+,0,j}$ and $\tilde{\lambda}_{+,0,j}$ and a tower of Dirac fermions $\Psi_{\pm,n,j}$ with masses
\begin{align}
m^2_{\Psi_{\pm,n,j}} = (-gf) (n+1) = \frac{4\pi |N|}{L^2} (n+1) \,.
\end{align}

Some comments are in order. The Abelian flux background is perturbatively stable, which means that all fields have a non-negative mass in the background field \eqref{symmetricgauge}. This situation is different for non-Abelian flux. The flux background can be associated with a gauge field in the Cartan subalgebra. The non-Cartan elements will accordingly be charged under the flux and some of the extra dimensional gauge field components become tachyonic. Therefore, the effective action below does not correspond to an expansion around the ground state of the system but rather around an extremal point. Nevertheless, it might be very interesting to study tachyon condensation in this framework and its interplay with the internal flux background. The study of tachyon condensation should reveal the true ground state of the theory, and the properties of the theory in the ground state could then be studied by shifting the vacuum accordingly, with possible applications to string theory \cite{Hashimoto:2003xz,Sen:2004nf}.

\section{Quantum corrections}
\label{sec:Radcorr}

In the previous sections we have derived four-dimensional effective actions for six-dimensional gauge theories compactified on a torus without or with magnetic flux, keeping the complete tower of charged excitations. This is a good starting point for computing quantum corrections, in particular for scalar masses which generically are not protected by symmetries. In the case without flux the one-loop effective potential of a Wilson line has been computed, and after subtraction of a divergent contribution a finite mass squared is obtained which is proportional to the inverse volume of the compact dimensions \cite{Antoniadis:2001cv, Cheng:2002iz, Ghilencea:2005hm, Ghilencea:2005nt}. Our main interest concerns quantum corrections to the Wilson line mass for a torus compactification with magnetic flux, but for comparison we first reconsider the case without flux.

\subsection{Quantum corrections without flux}
\label{subsec:noflux}
The one-loop corrections to the Wilson line mass are determined by the couplings of $\varphi$ to the matter fields $\Q$ and $\tilde\Q$. Gauge field contributions only enter at two-loop level. Hence our starting point is the action \eqref{S4-0} from which one obtains the Lagrangian in component form,
\begin{equation}
\begin{split}
\mathcal{L}_4 \supset &- \frac{1}{4} F_{\mu \nu} F^{\mu \nu} -  \partial_{\mu} \overline{\varphi} \partial^{\mu} \varphi  \\
& +\sum_{n,m} \big( -D_{\mu} \overline{\Q}_{n,m} D^{\mu} \Q_{n,m} +
  |M_{n,m} + \sqrt{2} gq \varphi|^2\ \overline{\Q}_{n,m} \Q_{n,m} \\
&\hspace{1.5cm} -i \chi_{n,m} \sigma^{\mu} D_{\mu} \overline{\chi}_{n,m}  -i \tilde{\chi}_{n,m} \sigma^{\mu} D_{\mu} \overline{\tilde{\chi}}_{n,m} \\
&\hspace{1.5cm} + (M_{n,m} + \sqrt{2} gq \varphi)\ \tilde{\chi}_{n,m}
\chi_{n,m} + \text{h.c.}\big)\,, 
\label{effectivenoflux}
\end{split}
\end{equation}
where the complex mass terms $M_{n,m}$ are defined in Eq.~\eqref{Mnm} and $D_\mu = \partial_\mu + igq A_{\mu}$.

Given the effective action \eqref{effectivenoflux} it is straightforward to calculate the one-loop quantum corrections to the Wilson line mass. The relevant bosonic and fermionic contributions are depicted in Fig.~\ref{noflux1loopb} and Fig.~\ref{noflux1loopf}, respectively, from which one obtains after a Wick rotation
\begin{figure}[t]
\centering
	\subfloat{
		\begin{overpic}[width= 0.20\textwidth]{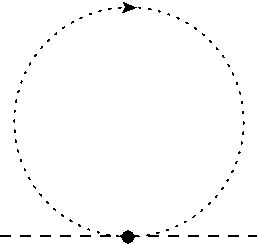}
			\put(-10, 0){$\varphi$}
			\put(102, 0){$\overline{\varphi}$}
			\put(35, 73){$\Q_{n,m}$}
		\end{overpic}
	}
	\hspace{0.15 \textwidth}
	\subfloat{
		\begin{overpic}[width=0.31 \textwidth]{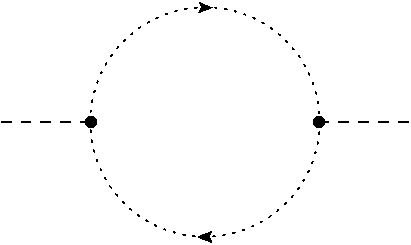}
			\put(-8, 28){$\varphi$}
			\put(102, 28){$\overline{\varphi}$}
			\put(40, 64){$\Q_{n,m}$}
			\put(40, 9){$\Q_{n,m}$}
		\end{overpic}
	}
\vspace{0.2cm}	
\caption{Bosonic contributions to the Wilson line mass without flux.}
	\label{noflux1loopb}
\vspace*{1.2cm}
\centering
	\begin{overpic}[width= 0.31\textwidth]{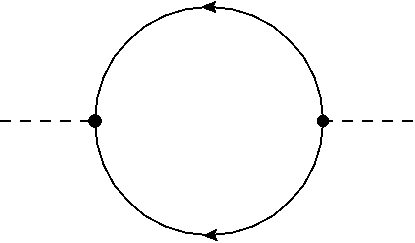}
		\put(-8, 28){$\varphi$}
		\put(102, 28){$\overline{\varphi}$}
		\put(46, 63){$\chi_{n,m}$}
		\put(46, 11){$\tilde{\chi}_{n,m}$}
	\end{overpic}
\vspace{0.2cm}	
\caption{Fermionic contribution to the Wilson line mass without flux.}
	\label{noflux1loopf}
\end{figure}
\begin{align}
\delta m_b^2 =
2g^2q^2 \sum_{n,m}  \int \frac{d^4 k}{(2\pi)^4} \left( \frac{1}{k^2 + |M_{n,m}|^2} - \frac{M_{n,m}\overline{M}_{n,m}}{(k^2 + |M_{n,m}|^2)^2}\right) \,,
\label{nofluxbos}
\end{align}
and
\begin{align}
\delta m_f^2  = -4 g^2q^2 \sum_{n,m}  \int \frac{d^4 k}{(2\pi)^4} \frac{k^2}{(k^2 + |M_{n,m}|^2)^2} = - 2 \delta m_b^2 \,.
\end{align}
As expected the contributions cancel for a supersymmetric spectrum, i.e.~for two charged scalars and a pair of charged Weyl fermions with the same masses.

The bosonic contribution \eqref{nofluxbos} to the Wilson line mass is an infinite sum of quadratically divergent terms. A consistent treatment of this expression requires a regularization prescription as well as renormalization conditions. In the literature several approaches have been pursued which make use of string inspired Poisson resummation \cite{Antoniadis:2001cv, Cheng:2002iz, Ghilencea:2001bv}, as well as dimensional regularization \cite{Ghilencea:2005vm}. In the following, we shall adopt the treatment in \cite{Antoniadis:2001cv}, which yields a well-known result for the Wilson line effective potential.

Using the Schwinger representation it is conveniently expressed as 
\begin{equation}
\begin{split}
\delta m_b^2 =& 2 g^2q^2 \sum_{n,m}\int_0^{\infty} dt \, t\, e^{- |M_{n,m}|^2 t} \int \frac{d^4 k}{(2\pi)^4} k^2 e^{- k^2 t} \\
=& \frac{g^2q^2}{4 \pi^2} \int_0^{\infty} \frac{dt}{t^2} \enspace \Theta_3\!\left(0; \frac{4 \pi i t}{L^2}\right)^2 \,,
\end{split}
\end{equation}
where we have interchanged summation over KK modes and t-integration, so that the integrand is now given by the square of the Jacobi $\Theta$-function
\begin{equation}
\Theta_3(z ; \tau) = \sum_r e^{i \pi \tau r^2} e^{2\pi i z r} \,. \label{Theta3}
\end{equation}
Under modular transformations $\Theta_3$ transforms as
\begin{align}
\Theta_3 (0; \tau) = (-i \tau)^{-1/2} \Theta_3 (0; - 1/\tau) \,.
\end{align}
From this we obtain
\begin{equation}
\begin{split}
\delta m_b^2 =&\frac{g^2q^2 L^2}{16 \pi^3} \int_0^{\infty} \frac{dt}{t^3}\ \Theta_3\!\left(0; \frac{i L^2}{4 \pi t}\right)^2 \\ 
=& \frac{g^2q^2 L^2}{16 \pi^3} \int_0^{\infty} du \, u \, \Theta_3\!\left(0; \frac{i L^2 u}{4 \pi}\right)^2 \\
=& \frac{g^2q^2}{\pi^3 L^2} \sum_{r,s} \frac{1}{(r^2 + s^2)^2}\,, \label{divmass}
\end{split}
\end{equation}
where we have used the explicit form \eqref{Theta3} in the last step.

The full divergence of $\delta m^2_b$, i.e.~summation over KK modes and quadratically divergent momentum integrations, is contained in the $r=s=0$ contribution to the sum in Eq.~\eqref{divmass}. To remove this divergence a counterterm is needed. Following \cite{Antoniadis:2001cv, Cheng:2002iz}, we define the finite part of $\delta m_b^2$ by dropping the $r=s=0$ contribution to the sum \eqref{divmass}. We have compared this finite part with the result in \cite{Ghilencea:2005vm}, which has been obtained by using dimensional regularization and Poisson resummation. It is reassuring that both procedures give the same answer.

The mass of $\varphi$ can also be obtained from the second derivative of the Wilson line effective potential which was calculated in \cite{Antoniadis:2001cv}. Here one starts from the effective mass of the matter fields $\Q_{n,m}$ in a constant Wilson line background $\varphi$ (see Eq.~\eqref{effectivenoflux}),
\begin{equation}
M_{n,m} (\varphi) = M_{n,m} + \sqrt{2}gq\varphi \,.
\end{equation}
From the general expression for the one-loop effective potential,
\begin{equation}
\begin{split}
V_{\text{eff}} &= \frac{1}{2} \sum_I (-1)^{F_I} \int \frac{d^4 k}{(2\pi)^4} \log \left( k^2 + M_I^2(\varphi) \right) \\
&= - \frac{1}{32 \pi^2} \sum_I (-1)^{F_I} \int \frac{dt}{t^3} \enspace e^{- M_I^2 (\varphi) t} \,,
\end{split}
\end{equation}
with $I$ labeling the bosons and fermions in the theory,  one obtains for the contribution of the complete KK tower of a single 6d charged scalar field $\Q$,
\begin{equation}
\begin{split}
V_{\text{eff}} &= - \frac{1}{16 \pi^2} \sum_{n,m} \int \frac{dt}{t^3} \enspace e^{- |M_{n,m}|^2 (\varphi) t} \\
&=  - \frac{1}{16 \pi^2} \sum_{n,m} \int_0^{\infty} \frac{dt}{t^3} \enspace \exp \left[ - \left( \frac{2\pi n}{L} + q g a_5 \right)^2 t - \left( \frac{2\pi m}{L} + q g a_6 \right)^2 t \right] \,,
\end{split}
\end{equation}
where $a_5$ and $a_6$ are constant background fields. After a Poisson resummation one finds
\begin{equation}
\begin{split}
V_{\text{eff}}
&= - \frac{L^2}{64 \pi^3} \sum_{r,s} \int_0^{\infty} \frac{dt}{t^4} \exp \left[ i q g L \left( r a_5 + s a_6\right) - \frac{L^2}{4t} \left(r^2 + s^2\right)\right] \\
&= - \frac{L^2}{64\pi^3} \sum_{r,s} \int_0^{\infty} du \, u^2 \exp \left[ i q g L \left( r a_5 + s a_6\right) - \frac{L^2}{4} u \left( r^2 + s^2 \right)\right]\,.
\end{split}
\end{equation}
Performing the $u$-integration and expressing the effective potential in terms of $\varphi$ yields the final result
\begin{align}
V_{\text{eff}} = - \frac{2}{L^4 \pi^3} \sum_{r,s} \frac{1}{(r^2 + s^2)^3} \exp \left[ i \frac{q g L}{\sqrt{2}} \left((s - i r) \varphi + (s + i r) \overline{\varphi} \right)\right]\,.
\end{align}
The $r=s=0$ contribution is again divergent. It has been argued that dropping this term corresponds to subtracting a divergent cosmological constant. However, since the expression for the effective potential is divergent, omitting the $r=s=0$ contribution also subtracts field dependent terms. Indeed, the mass term
\begin{align}
\partial_{\varphi} \partial_{\overline{\varphi}} V_{\text{eff}} |_{\varphi = 0}  =  \frac{g^2 q^2}{L^2 \pi^3} \sum_{r,s} \frac{1}{\left( r^2 + s^2 \right)^2 }\,
\end{align}
is divergent and identical to the expression \eqref{divmass}. On the other hand, the prescription to drop the $r=s=0$ is consistent with respect to the finite contributions, since the finite mass terms obtained from the diagrammatic calculation and the effective potential calculation then yield the same result.

\subsection{Quantum corrections with flux}
\label{subsec:Radflux}

Given the four-dimensional effective action for the torus compactification with flux, see \eqref{effectiveaction} and \eqref{effectivecomponent}, containing the complete tower of Landau levels we can again study quantum corrections to the Wilson line effective potential. In the following we shall compute the quantum corrections to the mass term and the quartic coupling.

From Eq.~\eqref{effectivecomponent} one reads off the couplings of the Wilson line $\varphi$ to the towers of charged bosonic and fermionic fields,
\begin{equation}
\begin{split}
\mathcal{L}_{int} = & -i \sqrt{2} q g \sum_{n,j} \sqrt{\alpha (n+1)}\
\varphi \left( \overline{\tilde{\Q}}_{n+1,j} \tilde{\Q}_{n,j} - \overline{\Q}_{n,j} \Q_{n+1,j} \right) + \text{h.c.} \\
& - 2 q^2 g^2 \sum_{n,j} |\varphi|^2 \left( |\Q_{n,j}|^2 + |\tilde{\Q}_{n,j}|^2 \right) \\
&- \sqrt{2} q g \sum_{n,j} \varphi \, \tilde{\chi}_{n,j} \chi_{n,j} + \text{h.c.} \,,
\label{Lint}
\end{split}
\end{equation}
where we have introduced the positive parameter $\alpha = -2qgf$ of mass dimension two. Note that the cubic bosonic vertex is proportional to the mass of the charged fields involved. Moreover, the bosonic couplings do not mix the fields $\Q$ and $\tilde{\Q}$. On the contrary, the fermionic coupling  involves the pair $\chi$ and $\tilde{\chi}$ at the same Landau level $n$, analogously to the Dirac mass terms in Eq.\ \eqref{massL}.

As in the case without flux there are two classes of bosonic contributions and one class of fermionic contributions to the Wilson line mass which are depicted in Fig.~\ref{phi2bos} and Fig.~\ref{phi2fer}, respectively.
\begin{figure}[t]
\centering
	\subfloat{
		\begin{overpic}[width= 0.20\textwidth]{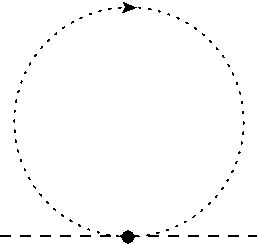}
			\put(-10, 0){$\varphi$}
			\put(102, 0){$\overline{\varphi}$}
			\put(25, 73){$\Q_{n,j}, \tilde{\Q}_{n,j}$}
		\end{overpic}
	}
	\hspace{0.15 \textwidth}
	\subfloat{
		\begin{overpic}[width=0.31 \textwidth]{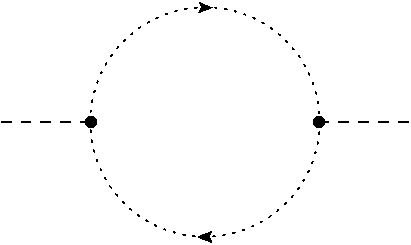}
			\put(-8, 28){$\varphi$}
			\put(102, 28){$\overline{\varphi}$}
			\put(25, 63){$\Q_{n+1,j},\tilde{\Q}_{n+1,j}$}
			\put(33, 11){$\Q_{n,j},\tilde{\Q}_{n,j}$}
		\end{overpic}
	}
\vspace{0.3cm}	
\caption{Bosonic contributions to the Wilson line mass with flux.}
	\label{phi2bos}
\vspace*{1.2cm}
\centering
	\begin{overpic}[width= 0.31\textwidth]{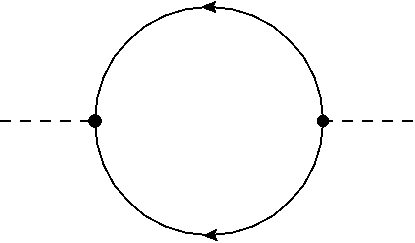}
		\put(-8, 28){$\varphi$}
		\put(102, 28){$\overline{\varphi}$}
		\put(46, 63){$\chi_{n,j}$}
		\put(46, 8){$\tilde{\chi}_{n,j}$}
	\end{overpic}
\vspace{0.3cm}	
\caption{Fermionic contribution to the Wilson line mass with flux.}
	\label{phi2fer}
\end{figure}
Using the couplings given in the Lagrangian \eqref{Lint} one obtains for the quantum corrections
\begin{equation}
\begin{split}
\delta m_b^2  &= 2 q^2 g^2 |N| \sum_{n} \int \frac{d^4k}{(2 \pi)^4} 
\left( \frac{2}{k^2 + \alpha (n + \tfrac{1}{2})} \right. \\
 & \left. \hspace{2.5cm} - \frac{2 \alpha (n+1)}{\left(k^2 + \alpha (n + \tfrac{3}{2})\right) \left(k^2 + \alpha (n + \tfrac{1}{2})\right)} \right) \,,\\
\delta m_f^2 &= -2 q^2 g^2 |N| \sum_n \int \frac{d^4k}{(2\pi)^4} \frac{2 k^2}{\left(k^2 + \alpha n \right)\left(k^2 + \alpha (n + 1)\right)} \,,
\end{split}
\end{equation}
which can be brought to the form
\begin{equation}
\begin{split}
\delta m^2_b &= -4 q^2 g^2 |N| \sum_n \int \frac{d^4 k}{(2 \pi)^4} \left( \frac{n}{k^2 + \alpha (n+\tfrac{1}{2})} - \frac{n+1}{k^2 + \alpha (n + \tfrac{3}{2})}\right) \,, \\
\delta m^2_f &= 4 q^2 g^2 |N| \sum_n \int \frac{d^4 k}{(2 \pi)^4} \left( \frac{n}{k^2 + \alpha n} - \frac{n+1}{k^2 + \alpha (n + 1)}\right) \,.
\end{split}
\end{equation}
Using the Schwinger representation of the propagators and performing the momentum integrations one finds
\begin{equation}
\begin{split}
\delta m^2_b  &= -\frac{q^2 g^2}{4 \pi^2} |N| \sum_n \int_0^{\infty} dt \, \frac{1}{t^2} \left( n e^{- \alpha (n+ \ttfrac{1}{2}) t} - (n+1) e^{- \alpha (n + \ttfrac{3}{2}) t}\right) \,, \\
\delta m^2_f  &= \frac{q^2 g^2}{4 \pi^2} |N| \sum_n \int_0^{\infty} dt \, \frac{1}{t^2} \left( n e^{- \alpha n t} - (n+1) e^{- \alpha (n + 1) t}\right) \,.
\end{split}
\end{equation}
As in the case without flux the bosonic as well as the fermionic contribution of each Landau level is quadratically divergent. However, interchanging summation and $t$-integration and using various identities for geometrical series, one arrives at
\begin{equation}
\begin{split}
\delta m^2_b =& -\frac{q^2 g^2}{4 \pi^2} |N| \int_0^{\infty} dt \, \frac{1}{t^2} \left( \frac{e^{\ttfrac{1}{2} \alpha t}}{(e^{\alpha t} - 1)^2} - \frac{e^{\ttfrac{1}{2} \alpha t}}{(e^{\alpha t} - 1)^2} \right) \\
=& \,0 \,, \\
\end{split}
\end{equation}
\begin{equation}
\begin{split}
\delta m^2_f  =&  \frac{q^2 g^2}{4 \pi^2} |N| \int_0^{\infty} dt \, \frac{1}{t^2} \left( \frac{e^{\alpha t}}{(e^{\alpha t} - 1)^2} - \frac{e^{\alpha t}}{(e^{\alpha t} - 1)^2} \right) \\
=& \, 0 \,.
\end{split}
\end{equation}
We conclude that, contrary to the case without flux, the contributions from the different Landau levels add up to zero and the integrand vanishes. It is remarkable that the bosonic and the fermionic contribution to the Wilson line mass vanish individually. To obtain this result it is important to perform the summation before the momentum integration, as in \cite{Antoniadis:2001cv, Ghilencea:2005vm}. In this way, the symmetries of the gauge theory in the compact dimensions are kept. Comparing the result with the case without flux suggests that  magnetic flux may provide a protection of the Wilson line mass compared to the compactification scale, independent of supersymmetry.

With non-vanishing flux the computation of the complete Wilson line effective potential is not straightforward, unlike in the case without flux. As next step we therefore compute the one-loop contribution to the quartic coupling $\lambda$. The calculation is very similar to the one for the mass term, although more cumbersome. The diagrams with charged fermions and bosons in the loops are depicted in Fig.~\ref{phi4fer} and Fig.~\ref{phi4bos}, respectively. Compared to the computation of the mass term now also fermion propagators appear that mix neighboring Landau levels. 
\begin{figure}[t]
\centering
	\subfloat{
		\begin{overpic}[width= 0.25\textwidth]{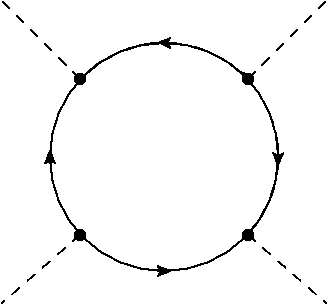}
			\put(-8, 90){$\varphi$}
			\put(-8, 0){$\overline{\varphi}$}
			\put(102, 90){$\overline{\varphi}$}
			\put(102, 0){$\varphi$}
			\put(46, 85){$\tilde{\chi}_{n,j}$}
			\put(46, 0){$\tilde{\chi}_{n,j}$}
			\put(-4, 44){$\chi_{n,j}$}
			\put(87, 44){$\chi_{n,j}$}
		\end{overpic}
	}
	\hspace{0.08 \textwidth}
	\subfloat{
		\begin{overpic}[width=0.25 \textwidth]{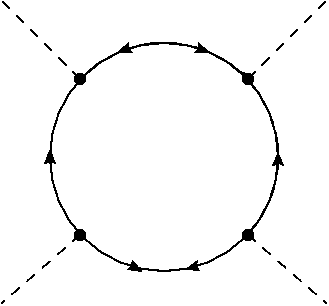}
			\put(-8, 90){$\varphi$}
			\put(-8, 0){$\overline{\varphi}$}
			\put(102, 90){$\varphi$}
			\put(102, 0){$\overline{\varphi}$}
			\put(-4, 44){$\chi_{n,j}$}
			\put(87, 44){$\tilde{\chi}_{n-1,j}$}
			\put(28, 84){$\tilde{\chi}_{n,j}$}
			\put(52, 84){$\chi_{n-1,j}$}
			\put(28, 3){$\tilde{\chi}_{n,j}$}
			\put(52, 3){$\chi_{n-1,j}$}
		\end{overpic}
	}
	\hspace{0.08 \textwidth}
	\subfloat{
		\begin{overpic}[width= 0.25\textwidth]{./4fermion23.png}
			\put(-8, 90){$\varphi$}
			\put(-8, 0){$\overline{\varphi}$}
			\put(102, 90){$\varphi$}
			\put(102, 0){$\overline{\varphi}$}
			\put(-4, 44){$\tilde{\chi}_{n,j}$}
			\put(87, 44){$\chi_{n+1,j}$}
			\put(28, 84){$\chi_{n,j}$}
			\put(52, 84){$\tilde{\chi}_{n+1,j}$}
			\put(28, 3){$\chi_{n,j}$}
			\put(52, 3){$\tilde{\chi}_{n+1, j}$}
		\end{overpic}
	}
\vspace{0.3cm}	
\caption{Fermionic contributions to the Wilson line quartic
          coupling with flux.}
	\label{phi4fer}
\end{figure}
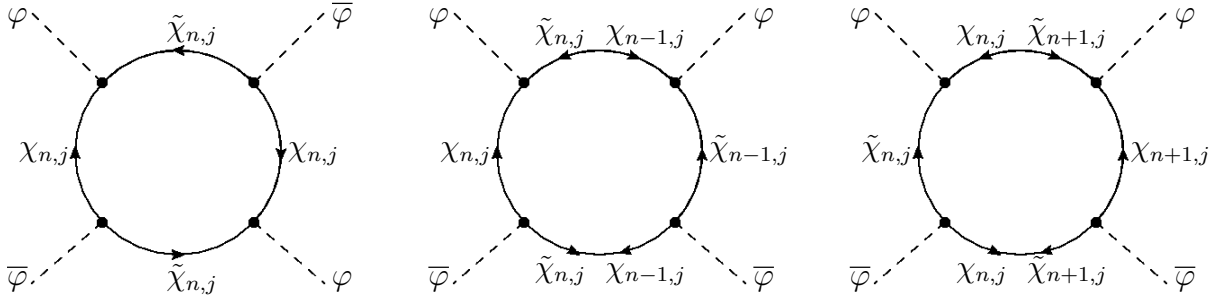
As for the mass term we calculate the contributions from bosons and fermions separately.
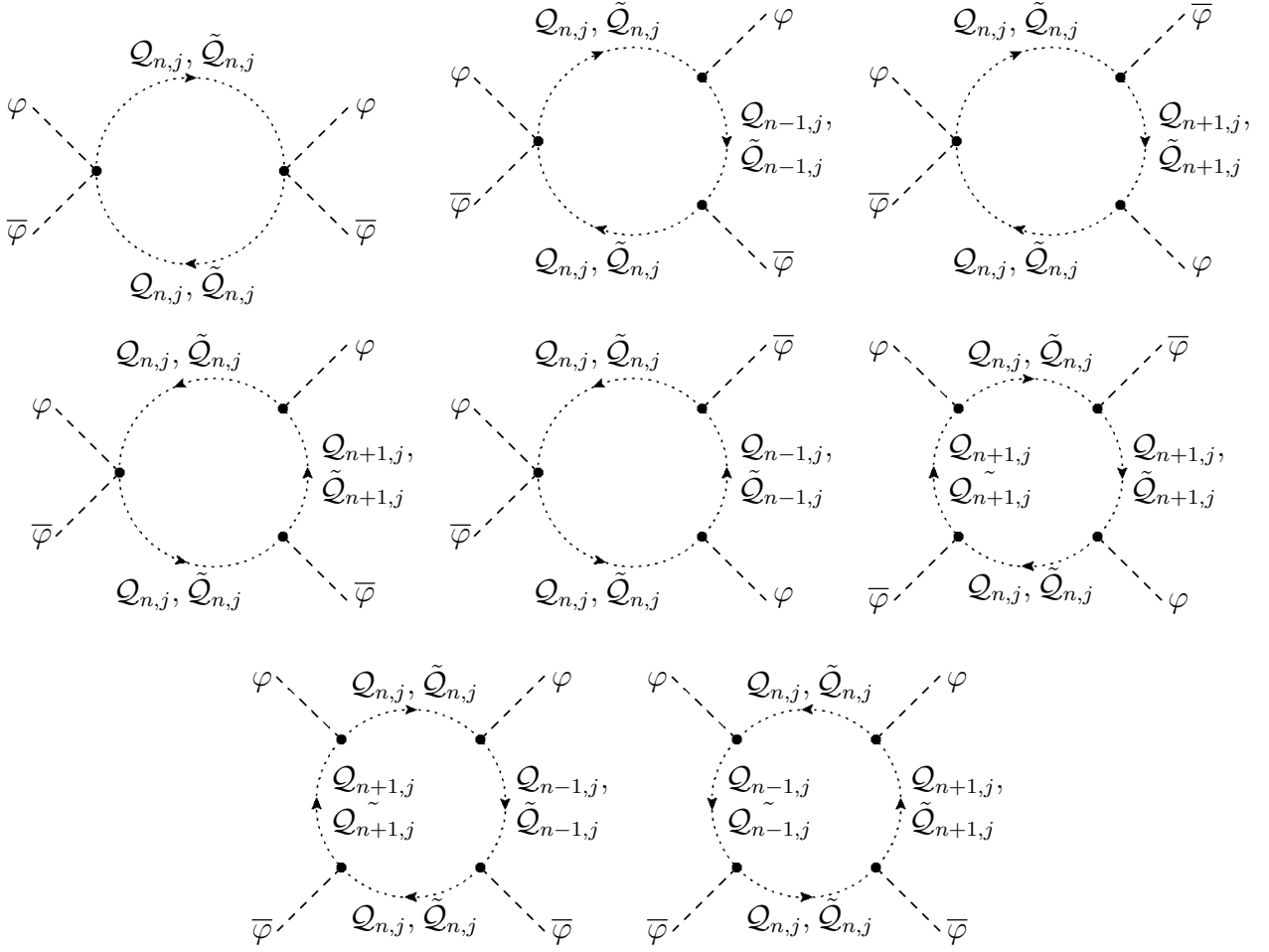
\begin{figure}[t]
\centering
	\subfloat{
		\begin{overpic}[width= 0.27\textwidth]{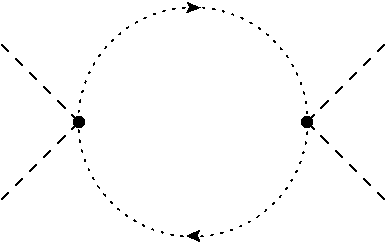}
			\put(-8, 50){$\varphi$}
			\put(-8, 10){$\overline{\varphi}$}
			\put(102, 50){$\varphi$}
			\put(102, 10){$\overline{\varphi}$}
			\put(30, 66){$\Q_{n,j},\tilde{\Q}_{n,j}$}
			\put(30, -8){$\Q_{n,j},\tilde{\Q}_{n,j}$}
		\end{overpic}
	}
	\hspace{0.08 \textwidth}
	\subfloat{
		\begin{overpic}[width=0.25 \textwidth]{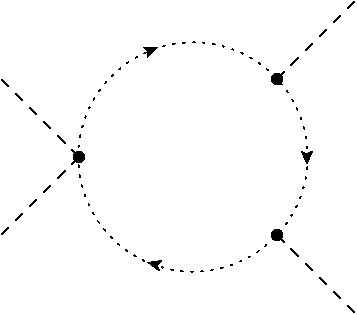}
			\put(-8, 65){$\varphi$}
			\put(-8, 20){$\overline{\varphi}$}
			\put(102, 85){$\varphi$}
			\put(102, 0){$\overline{\varphi}$}
			\put(20, 82){$\Q_{n,j},\tilde{\Q}_{n,j}$}
			\put(20, 0){$\Q_{n,j},\tilde{\Q}_{n,j}$}
			\put(90, 50){$\Q_{n-1,j},$}
			\put(90, 35){$\tilde{\Q}_{n-1,j}$}
		\end{overpic}
	}
	\hspace{0.08 \textwidth}
	\subfloat{
		\begin{overpic}[width= 0.25\textwidth]{./4boson23.png}
			\put(-8, 65){$\varphi$}
			\put(-8, 20){$\overline{\varphi}$}
			\put(102, 85){$\overline{\varphi}$}
			\put(102, 0){$\varphi$}
			\put(20, 82){$\Q_{n,j},\tilde{\Q}_{n,j}$}
			\put(20, 0){$\Q_{n,j},\tilde{\Q}_{n,j}$}
			\put(90, 50){$\Q_{n+1,j},$}
			\put(90, 35){$\tilde{\Q}_{n+1,j}$}
		\end{overpic}
	} \\ \vspace{0.6cm}
	\subfloat{
		\begin{overpic}[width= 0.25\textwidth]{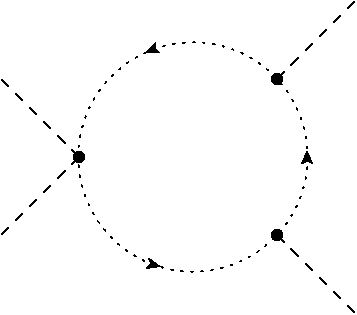}
			\put(-8, 65){$\varphi$}
			\put(-8, 20){$\overline{\varphi}$}
			\put(102, 85){$\varphi$}
			\put(102, 0){$\overline{\varphi}$}
			\put(20, 82){$\Q_{n,j},\tilde{\Q}_{n,j}$}
			\put(20, 0){$\Q_{n,j},\tilde{\Q}_{n,j}$}
			\put(90, 50){$\Q_{n+1,j},$}
			\put(90, 35){$\tilde{\Q}_{n+1,j}$}
		\end{overpic}
	}
	\hspace{0.08 \textwidth}
	\subfloat{
		\begin{overpic}[width=0.25 \textwidth]{./4boson45.png}
			\put(-8, 65){$\varphi$}
			\put(-8, 20){$\overline{\varphi}$}
			\put(102, 85){$\overline{\varphi}$}
			\put(102, 0){$\varphi$}
			\put(20, 82){$\Q_{n,j},\tilde{\Q}_{n,j}$}
			\put(20, 0){$\Q_{n,j},\tilde{\Q}_{n,j}$}
			\put(90, 50){$\Q_{n-1,j},$}
			\put(90, 35){$\tilde{\Q}_{n-1,j}$}
		\end{overpic}
	}
	\hspace{0.08 \textwidth}
	\subfloat{
		\begin{overpic}[width= 0.23\textwidth]{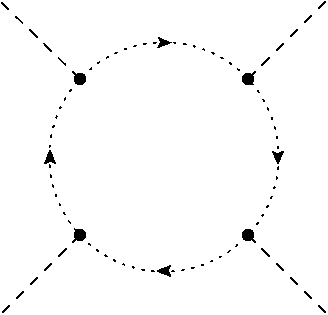}
			\put(-9, 92){$\varphi$}
			\put(-9, -2){$\overline{\varphi}$}
			\put(102, 92){$\overline{\varphi}$}
			\put(102, -2){$\varphi$}
			\put(27, 89){$\Q_{n,j},\tilde{\Q}_{n,j}$}
			\put(27, 2){$\Q_{n,j},\tilde{\Q}_{n,j}$}
			\put(88, 54){$\Q_{n+1,j},$}
			\put(88, 38){$\tilde{\Q}_{n+1,j}$}
			\put(20,54){$\Q_{n+1,j}$}
			\put(20,38){$\tilde{\Q_{n+1,j}}$}
		\end{overpic}
	} \\ \vspace{0.6cm}
	\subfloat{
		\begin{overpic}[width= 0.23\textwidth]{./4boson67.png}
			\put(-9, 92){$\varphi$}
			\put(-9, -2){$\overline{\varphi}$}
			\put(102, 92){$\varphi$}
			\put(102, -2){$\overline{\varphi}$}
			\put(27, 89){$\Q_{n,j},\tilde{\Q}_{n,j}$}
			\put(27, 2){$\Q_{n,j},\tilde{\Q}_{n,j}$}
			\put(88, 54){$\Q_{n-1,j},$}
			\put(88, 38){$\tilde{\Q}_{n-1,j}$}
			\put(20,54){$\Q_{n+1,j}$}
			\put(20,38){$\tilde{\Q_{n+1,j}}$}
		\end{overpic}
	}
	\hspace{0.08 \textwidth}
	\subfloat{
		\begin{overpic}[width=0.23 \textwidth]{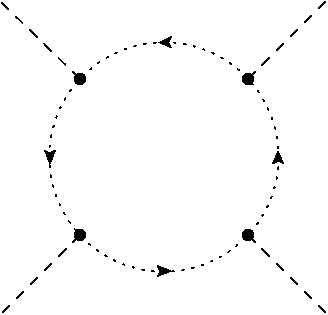}
			\put(-9, 92){$\varphi$}
			\put(-9, -2){$\overline{\varphi}$}
			\put(102, 92){$\varphi$}
			\put(102, -2){$\overline{\varphi}$}
			\put(27, 89){$\Q_{n,j},\tilde{\Q}_{n,j}$}
			\put(27, 2){$\Q_{n,j},\tilde{\Q}_{n,j}$}
			\put(88, 54){$\Q_{n+1,j},$}
			\put(88, 38){$\tilde{\Q}_{n+1,j}$}
			\put(20,54){$\Q_{n-1,j}$}
			\put(20,38){$\tilde{\Q_{n-1,j}}$}
		\end{overpic}
	}
\vspace{0.3cm}	
\caption{Bosonic contributions to the Wilson line quartic coupling with flux.}
\label{phi4bos}
\end{figure}
After some manipulations of the integrand one obtains the result
\begin{equation}
\begin{split}
 \delta \lambda_b = & -8 q^4 g^4 |N| \sum_n \int \frac{d^4k}{(2 \pi)^4} \left[ \frac{n^2}{A_{1/2}^2} - \frac{(n+1)^2}{A_{3/2}^2} \right.  \\
& \hspace{0.7cm} \left. + \frac{1}{\alpha} \left( - \frac{n(n+1)}{A_{1/2}} + \frac{(n+1)(n+2)}{A_{3/2}} + \frac{n(n+1)}{A_{3/2}} - \frac{(n+1)(n+2)}{A_{5/2}} \right) \right] \,, \\
\delta \lambda_f = & -8 q^4 g^4 |N| \sum_n \int \frac{d^4k}{(2 \pi)^4} \left[ - \frac{n^2}{(A_0)^2} + \frac{(n+1)^2}{(A_1)^2}\right. \\
& \hspace{0.7cm} \left. + \frac{1}{\alpha} \left( \frac{n(n+1)}{A_0} - \frac{(n+1)(n+2)}{A_1} - \frac{n(n+1)}{A_1} + \frac{(n+1)(n+2)}{A_2}\right)\right] \,,
\end{split}
\end{equation}
where we have introduced the shorthand notation $A_j = k^2 + \alpha(n+j)$. Introducing again the Schwinger representation of the propagators, performing the momentum integrations and interchanging summation and t-integration yields
\begin{equation}\label{quarticb}
\begin{split}
\delta \lambda_b  &=  -\frac{q^4 g^4}{2 \pi^2} |N| \int_0^{\infty} dt \sum_n\left(-e^{-\ttfrac{1}{2} \alpha t}\right) \Big[ \frac{1}{t} \left( -n^2 e^{-\alpha n t} + (n+1)^2 e^{-\alpha (n+1) t}\right) \\
&\hspace{3cm} + \frac{1}{\alpha t^2} \big(n(n+1) e^{-\alpha n t} - (n+1)(n+2) e^{- \alpha (n+1) t} \\
&\hspace{3cm} - n(n+1) e^{-\alpha (n+1) t} + (n+1)(n+2) e^{-\alpha (n+2) t}\big)\Big] \\
& = \, 0 \,, 
\end{split}
\end{equation}
\begin{equation}\label{quarticf}
\begin{split}
\delta \lambda_f &= -\frac{q^4 g^4}{2 \pi^2} |N| \int_0^{\infty} dt \sum_n\Big[ \frac{1}{t} \left( -n^2 e^{-\alpha n t} + (n+1)^2 e^{-\alpha (n+1) t}\right) \\
&\hspace{3cm} + \frac{1}{\alpha t^2} \big( n(n+1) e^{-\alpha n t} -
     (n+1)(n+2)e^{- \alpha (n+1) t} \\
&\hspace{3cm} - n(n+1) e^{-\alpha (n+1) t} + (n+1)(n+2) e^{-\alpha
      (n+2) t}\big)\Big] \,.\\
& = \, 0 \,.
\end{split}
\end{equation}
The sums in Eqs.~\eqref{quarticb} and \eqref{quarticf} extend from $0$ to $+\infty$. Since they are convergent and the $n=0$ contribution vanishes one can perform a shift $n \rightarrow n+1$ in the first term of each line. It is then apparent that the bosonic and the fermionic contribution to the quartic coupling again vanish separately. Hence, no $|\varphi|^4$-term is generated at one-loop order. This suggests that the entire one-loop effective potential vanishes. Indeed, this has already been conjectured in the original paper by Bachas \cite{Bachas:1995ik} based on the independence of the Landau level masses on the the Wilson lines. At the level of the 4d effective action this result appears very surprizing but, as we shall see in the following section, it can be understood in terms of symmetries of the six-dimensional theory. 

\section{Wilson lines as Goldstone bosons}
\label{sec:goldstone}

From the 4d effective field theory perspective the vanishing of the quantum corrections to the Wilson line effective potential is far from obvious. It is a consequence of an intricate interplay between level-dependent masses and couplings. Furthermore, the separate cancellations in the bosonic and fermionic sectors show that also supersymmetry is not responsible for this protection of scalar masses by magnetic flux. Considering the 6d theory it becomes clear which symmetry lies behind the vanishing of the effective potential for $\varphi$. The massless Wilson lines are the Goldstone bosons of the translation symmetries that are spontaneously broken by the background gauge field. We subsequently analyze the cases with a single $U(1)$ gauge group and with several $U(1)$ factors.

\subsection{Goldstone bosons for a single U(1)}
\label{subsec:symnoflux}

The six-dimensional action of a charged matter field that we considered in the previous sections,
\begin{align}
S_6 = \int d^6 x \, \left( - D_M \Qb D^M \Q \right) \,, \nonumber
\label{6dboson}
\end{align}
with $D_M \Q = (\partial_M + i q g \, A_M) \Q$, is obviously invariant under translations in the two torus directions,
\begin{align}
\delta \Q = \epsilon^m \partial_m \Q \,, \quad \delta A_n =
\epsilon^m \partial_m A_n \,,
\end{align}
implying $\delta D_M\Q = \epsilon^m \partial_m D_M\Q $ and therefore $\delta S_6 = 0$. In Sec.~\ref{sec:Abelianflux} we considered effective actions where the KK tower of the gauge field was neglected, i.e., the gauge field was replaced by its zero-mode, the complex Wilson line $\varphi = \tfrac{1}{\sqrt{2}}\left( a_6 + i a_5\right)$. The corresponding 6d action is invariant under the transformation
\begin{align}
\delta \Q = \epsilon^m \partial_m \Q \,, \quad \delta a_n = 0 \,. \label{tsym}
\end{align}
Let us now include magnetic flux by changing the covariant derivative to
\begin{align}
D_m \Q = \left(\partial_m + iq g\left(a_m + \tfrac{f}{2} \epsilon_{mn} x_n \right) \right) \Q \,.
\label{covflux}
\end{align}
The background gauge field $\langle A_m\rangle = \tfrac{f}{2} \epsilon_{mn} x_n$ breaks the translational $U(1)\times U(1)$ symmetry spontaneously. Now this symmetry is realized nonlinearly,
\begin{align}
\delta \Q = \epsilon^m \partial_m \Q \,, \quad \delta a_n = \epsilon^m \tfrac{f}{2} \epsilon_{nm} \,, \label{tsymnonl}
\end{align}
and the two real massless scalars $a_5$ and $a_6$ are the corresponding Goldstone bosons\footnote{There are other examples where the spontaneous breaking of translational invariance leads to the appearance of Goldstone bosons. For instance, the localization of a D$p$-brane in $9-p$ dimensions implies the existence of $9-p$ massless scalars localized on the D$p$-brane. See, for example, \cite{Ibanez:2012zz}.}. Note that the Wilson lines $a_5$ and $a_6$ remain massless if the KK tower $\hat{A}_m$ of massive scalars is included, i.e., $A_m = a_m + \hat{A}_m$, with the transformation behavior
\begin{align}
\delta \Q = \epsilon^m \partial_m \Q \,, \quad \delta a_n = \epsilon^m
\tfrac{f}{2} \epsilon_{nm}\,,\quad \delta \hat{A}_m = \epsilon^m \partial_m \hat{A}_m \,.
\label{tsymnonl2}
\end{align}
However, 6d gravity effects may modify the Wilson line masses, which remains to be investigated. In this connection also the backreaction of the flux on the geometry has to be taken into account.

The background gauge field used in Eq.~\eqref{covflux} corresponds to a particular choice of gauge. The same magnetic flux $F=d\langle A \rangle$ is generated by the background fields
\begin{align}
\langle A(x_5,x_6)\rangle = (a_5 - c f x_6) dx_5 + (a_6 + (1-c) f x_5) dx_6 \,,
\label{generalbackground}
\end{align}
with $c \in \mathbb{R}$. However, not all values of $c$ are allowed since the background gauge field has to satisfy the periodicity condition on a torus\footnote{For a recent discussion and references, see \cite{Buchmuller:2015eya}.},
\begin{equation}
\langle A(x_5 + k L, x_6 + l L) \rangle = \langle A(x_5, x_6)\rangle + d\Lambda \,, \quad k,l \in \mathbb{Z}\,, \label{period}
\end{equation}
where $\Lambda_{m,n}$ is a large gauge transformation,
\begin{equation}
\Lambda = \frac{2\pi}{L} (m x_5 + n x_6) \,, \quad m,n \in \mathbb{Z}\,. 
\end{equation}
This means that for all integers $k,l$ other integers $m,n$ have to exist such that the condition \eqref{period} is satisfied. Inserting the background field \eqref{generalbackground} into Eq.~\eqref{period} yields
\begin{equation}
-cflL dx_5 + (1-c)fkL dx_6 = \frac{2\pi}{L}(m dx_5 + n dx_6)\,,
\end{equation}
and with $fL^2/(2\pi) = N \in \mathbb{Z}$ this leads to the conditions $-cNl \in \mathbb{Z}$ and $(1-c)Nk \in \mathbb{Z}$, and therefore
\begin{equation}
cN \in \mathbb{Z}\,.
\end{equation}
For $c \neq 0$ and $c \neq 1$, it is apparent that the background gauge field \eqref{generalbackground} breaks both translational symmetries. For $c = 0$ or $1$ one might, at first sight, expect that one of the translations still is an unbroken symmetry, specifically, translations in $x_6$ for $c = 0$ and translations in $x_5$ for $c = 1$. However, in these cases the seemingly unbroken translational symmetry is broken by the periodicity condition. For instance, consider the case $c=1$. The background field $\langle A(x_5,x_6) \rangle = -f x_6 dx_5$ is changed by a torus translation to $\langle A(x_5+kL,x_6+lL) \rangle = -f x_6 dx_5 - flL dx_5 \equiv -f x_6 dx_5 + d\Lambda$, which yields
\begin{equation}
\Lambda = - \frac{2\pi}{L} Nl x_5\,.
\end{equation}
Clearly, the large gauge transformation that relates the two gauge fields connected by a torus translation breaks the translation symmetry in $x_5$-direction. We conclude that also for $c=1$ both translation symmetries are broken.  One easily confirms that the same is true in the case $c=0$.

\subsection{(Pseudo) Goldstone bosons for more U(1)'s}
\label{subsec:symlarge}

The situation becomes more subtle in the case of more than one $U(1)$ gauge group and an arbitrary number of charged scalars $\Q^i$ with different charge assignments. The covariant derivatives then read
\begin{align}
D_m \Q^i = \left( \partial_m + i q_{i \alpha} \left( a_m^{(\alpha)} + \tfrac{f^{(\alpha)}}{2} \epsilon_{mn} x_n \right)\right) \Q^i \,,
\end{align}
with $i$ and $\alpha$ labeling the various $U(1)$ gauge groups and charged matter fields, respectively.

In order to identify the Goldstone bosons, i.e. the nonlinearly transforming Wilson lines $a_m^{(\alpha)}$, we start from the individual translation symmetries for the charged fields,
\begin{align}
\delta \Q^i = \epsilon_{(i)}^m \partial_m \Q^i \,.
\label{moresym}
\end{align}
As in the previous section, the transformation behavior of the Wilson lines $a_m^{(\alpha)}$ is determined by the condition that the Lagrangian transforms into a total derivative, i.e. $\delta(D_m \Q^i) = \epsilon_{(i)}^m \partial_m D_m\Q^i$. This implies
\begin{align}
q_{i \alpha} \, \delta a_n^{(\alpha)} = q_{i \beta} \, \epsilon^m_{(i)} \, \tfrac{f^{(\beta)}}{2} \epsilon_{nm} \,,
\label{morenonlin}
\end{align}
for all $i$. This relation expresses the fact that, depending on the matrix $q_{i\alpha}$, fields charged under $U(1)_{\alpha}$ may feel an effective flux, even though the flux of the gauge group $U(1)_{\alpha}$ vanishes.
 
It is evident that for $N$ $U(1)$ gauge groups and $N_f$ charged fields there can be at most $\min (N, N_f)$ Goldstone bosons. If there is flux in at least one of the gauge groups the two translation symmetries are spontaneously broken and there are at least two Goldstone boson, as discussed in the previous section. Further symmetries are accidental in the sense that they may be explicitly broken by additional interactions that couple the various matter fields, such as $\left| \Q^i \right|^2 \left| \Q^j \right|^2$. Therefore, masses for some Wilson lines $a_m^{(\alpha)}$ may be generated beyond one-loop. 

In order to illustrate the subtleties in identifying the (pseudo) Goldstone bosons we discuss a simple example. Consider the gauge group $U(1)_1 \times U(1)_2$ and two matter fields $\Q^i$ with the charge matrix
\begin{align}
q_{i \alpha} = \begin{pmatrix} 1 & 1 \\ 1 & -1 \end{pmatrix} \,.
\end{align}
For vanishing fluxes, $f^{(1)} = f^{(2)} = 0$, it is obvious from the relation \eqref{morenonlin} that none of the fields $a_m^{(\alpha)}$ transforms nonlinearly. Hence, both symmetries \eqref{moresym} are preserved and there are no (pseudo) Goldstone bosons. For the flux assignment $f^{(1)} = f^{(2)} \equiv f\neq 0$ one finds
\begin{align}
q_{i \beta} f^{\beta} = f \begin{pmatrix} 2 \\ 0 \end{pmatrix} \,,
\end{align} 
and there is a single Goldstone boson corresponding to $a_m^{(1)} + a_m^{(2)}$. Finally, for $f^{(1)} \equiv f \neq 0$ and $f^{(2)} = 0$ one has
\begin{align}
q_{i \beta} f^{\beta} = f \begin{pmatrix} 1 \\ 1 \end{pmatrix} \,,
\end{align}
and both Wilson lines transform nonlinearly according to \eqref{morenonlin}, which corresponds to two (pseudo) Goldstone bosons.

\section{Conclusion and Outlook}
\label{sec:conclusion}

In this work we have derived the four-dimensional supersymmetric effective action for six-dimensional gauge theories compactified on a torus with various background gauge fields. For non-vanishing background flux we have shown how the Kaluza-Klein excitations of the vector multiplet obtain their masses from a supersymmetric St\"uckelberg mechanism, and we have determined their couplings to charged chiral multiplets. For non-vanishing flux in the internal dimensions we have restricted the uncharged sector to the zero modes. The entire tower of the charged states, however, is incorporated, and their modified mass spectrum is obtained by solving the $D$- and $F$-term equations. As is well known, the massive tower corresponds to a harmonic oscillator spectrum with helicity-dependent shifts, where each level is $|N|$-fold degenerate. The Abelian flux background is perturbatively stable and we have worked out the full effective action in superfields as well as components. For non-Abelian flux we have clarified the physical mass spectrum. The internal components of the gauge field develop a tachyonic direction, and we expect the derived supersymmetric effective action to prove useful for the treatment of tachyon condensation.

Using the explicit expressions for the level-dependent couplings of the charged tower to the Abelian Wilson lines, which are massless at tree-level, we have reproduced the known quantum corrections for vanishing flux. Following a regularization prescription used in the literature, one obtains a finite result. For non-vanishing flux the situation changes drastically. The quantum corrections induced by the charged bosons and fermions separately vanish at one-loop order for the $|\varphi|^2$ and $|\varphi|^4$ terms of the effective Lagrangian. This was shown using a diagrammatic approach where it follows from an intricate interplay between level-dependent masses and couplings. Considering the six-dimensional theory one understands that not just the  $|\varphi|^2$ and $|\varphi|^4$ terms, but the entire effective potential should vanish exactly. The background gauge field associated with the magnetic flux breaks the translation symmetry in the $x_5$- and $x_6$-directions spontaneously. This leads to two massless Goldstone bosons which can be identified as the Wilson lines $a_5$ and $a_6$ contained in the complex field $\varphi$.

The results described above suggest several extensions of our work. First of all, the analysis of globally supersymmetric gauge theories with magnetic flux should be extended to supergravity theories. This would allow to study the backreaction of the flux on the geometry of the compact dimensions as well as possible mixings between moduli of the metric and the Wilson lines. Very important are also flux compactifications on orbifolds, see e.g. \cite{Abe:2009uz, Buchmuller:2015eya,Kobayashi:2016qag, Abe:2016jsb}. In models with gauge-Higgs unification \cite{Hosotani:1983xw, Hatanaka:1998yp, Antoniadis:2001cv, Hall:2001pg, ArkaniHamed:2001nc} one could then investigate the effect of magnetic flux on quantum corrections to Higgs masses. It is an intriguing possibility that magnetic flux in higher dimensions may contribute significantly to stabilize the electroweak scale.

\section*{Acknowledgments}
We thank Fabian R\"uhle and Yannick Linke for valuable discussions. This work was supported by the German Science Foundation (DFG) within the Collaborative Research Center (SFB) 676 ``Particles, Strings and the Early Universe''. E.D. was supported in part by the ``Agence Nationale de la Recherche" (ANR). M.D. acknowledges support from the ``Studienstiftung des deutschen Volkes''.

}

\providecommand{\href}[2]{#2}\begingroup\raggedright\endgroup

\end{document}